\documentclass[conference]{IEEEtran}
\IEEEoverridecommandlockouts

\usepackage[utf8]{inputenc}
\usepackage[T1]{fontenc}
\usepackage{lipsum}
\usepackage{amsmath, amssymb, amsfonts}
\usepackage{graphicx}
\usepackage{booktabs}
\usepackage{multirow}
\usepackage{url}
\usepackage{xurl}
\usepackage[hidelinks]{hyperref}
\usepackage{xcolor}
\usepackage{cleveref}
\usepackage[colorinlistoftodos,textsize=footnotesize]{todonotes} 
\usepackage[table]{xcolor}
\usepackage{braket}

\definecolor{victimcolor}{RGB}{229,245,224}   
\definecolor{entcolor}{RGB}{199,230,255}      

\usepackage[
  defernumbers=true,
  backend=biber,
  sortcites=true,
  url=true,
  maxbibnames=10,
  maxcitenames=1,
  doi=false
]{biblatex}
\begin{document}

\title{An End-to-End Multi-Stage Kill-Chain Attack on Quantum Neural Networks: Demonstration on Trapped-Ion Hardware}

\author{
    \IEEEauthorblockN{
        Cedric Brügmann\IEEEauthorrefmark{1},
        Daniel Herr\IEEEauthorrefmark{1},
        Daniel Ohl de Mello\IEEEauthorrefmark{1},
        Pascal Debus\IEEEauthorrefmark{2}, \\
        Maximilian Wendlinger\IEEEauthorrefmark{2}, Kilian Tscharke\IEEEauthorrefmark{2}, Juris Ulmanis\IEEEauthorrefmark{3}, Alexander Erhard\IEEEauthorrefmark{3}, Arthur Schmidt\IEEEauthorrefmark{4}, Fabian Petsch\IEEEauthorrefmark{4}}
    \IEEEauthorblockA{
    \IEEEauthorrefmark{1}\textit{d-fine GmbH}, Frankfurt, Germany\\
    \IEEEauthorrefmark{2}\textit{Fraunhofer Institute for Applied and Integrated Security (AISEC)}, Garching near Munich, Germany\\
    \IEEEauthorrefmark{3}\textit{Alpine Quantum Technologies (AQT) GmbH}, Innsbruck, Austria\\
    \IEEEauthorrefmark{4}\textit{Federal Office for Information Security (BSI)}, Bonn, Germany
    }
}

\maketitle

\begin{abstract}
We demonstrate an end-to-end, multi-stage attack against a quantum neural network (QNN) model that is executed on a trapped-ion quantum computer. Our chain combines side-channel reconnaissance, crosstalk characterization, adversarial example generation, and a physical crosstalk attack that realizes the adversarial perturbation on the device. We cover the full attack chain on ion traps and report the corresponding superconducting-hardware experiments in the appendix. We discuss implications for QaaS providers and hardware mitigations.
\end{abstract}

\begin{IEEEkeywords}
Quantum machine learning, security, side-channel attacks, crosstalk, adversarial examples, trapped ions, superconducting qubits, kill-chain
\end{IEEEkeywords}

\section{Introduction}
Deployments of Quantum Machine Learning (QML) algorithms on real quantum processing units (QPUs) fundamentally expand the attack surface beyond what is captured by idealized or noise-augmented simulations. Existing work on the security of QML has primarily examined individual attack vectors in isolation. While these results provide valuable insights into specific vulnerabilities, they do not capture how an adversary may combine multiple techniques into a coherent attack campaign. In practical deployment scenarios, especially in cloud-based and multi-tenant settings, attacks are rarely executed as isolated events. Instead, information gained during reconnaissance can be used to reduce uncertainty in subsequent stages, enabling targeted manipulations. Building on an extensive review of the literature, we develop a structured taxonomy of QML attack vectors and align them with the respective stages of a quantum-aware kill chain, following the spirit of the MITRE ATLAS framework for classical ML \cite{debus_entangled_2025}. This framework makes explicit the interdependencies between different threat classes, spanning hardware-level weaknesses (e.g., side-channel leakage and crosstalk-induced faults), data and algorithm manipulation (e.g., poisoning and circuit backdoors), as well as privacy-focused attacks (e.g., model extraction and training data inference). By systematically evaluating these interdependencies, our approach facilitates the design of proactive and integrated defense-in-depth strategies. One of the findings of this kill chain perspective is the possibility of linked multi-stage attacks, where the attacker uses side-channel attacks to perform reconnaissance and learn as much as possible about a victim machine learning model, and then use this information to fine-tune attacks, e.g., based on noise injection, to attack a model. Importantly, all evaluated attack vectors operate within the constraints of current noisy intermediate-scale quantum (NISQ) devices and do not rely on fault-tolerant assumptions, making them directly relevant for near-term
quantum machine learning deployments. The remainder of this paper is organized as follows. Section \ref{sec:related_work} summarizes related work, positioning our contributions within the existing literature. Section \ref{sec:background} introduces the necessary background, covering both the theoretical foundations and the hardware-specific aspects. Section \ref{sec:methods} details the experimental pipeline, experimental setup, and attack methodologies employed in this study. Next, we present the results in Section \ref{sec:results}. These findings are subsequently interpreted and critically assessed in the discussion section \ref{sec:discussion}, where defensive strategies against the attack vectors are also discussed. Finally, in Section \ref{sec:conclusion}, we summarize the main contributions.

\section{Related Work}\label{sec:related_work}

\subsection{Side-Channel Attacks on Quantum Computers}\label{subsec:side_channel_attacks}
One of the main targets of side-channel attacks (SCA) is reverse engineering
of the quantum circuit. SCA based on power traces,
which can be calculated directly from the control electronics or
the pulse program, have already been described in the literature \cite{erata_quantum_2024, xu_exploration_2023}. In addition, timing-based approaches have been proposed in \cite{lu_quantum_2024} and successfully tested on IBM quantum computing hardware. Studies exploiting crosstalk as a side channel have also been reported  \cite{lee_swap_2025,choudhury_crosstalk-induced_2024}. However, all of these attacks were executed on superconducting devices, whereas our work focuses on trapped-ion platforms as well.


\subsection{Adversarial Attacks on QML}
Adversarial evasion attacks are characterized by the fact that they deliberately but minimally perturb the input of a (QML) model to achieve an incorrect classification by the model. 

\newpage A variety of different attacks have been researched in the classical case, and the vulnerability of QML methods was established (in theory) early on by Ref. \cite{liu_vulnerability_2020}. In empirical studies, vulnerability to standard white box
methods (Fast Gradient Sign Method  \cite{goodfellow_explaining_2015}, Projected Gradient
Descent \cite{madry_towards_2019}) was shown in various works in simulation \cite{lu_quantum_2020} and also on real quantum
hardware \cite{ren_experimental_2022} for both classical and quantum data. Simple defenses such as adversarial training, showing improved robustness under adversarial attack \cite{du_quantum_2021}. Ref. \cite{west_benchmarking_2023} also investigates transfer attacks as gray-box/black-box methods and suggests
a possible quantum advantage in robustness. Adversarial
robustness, as well as the influence of data encoding and
advanced defense mechanisms such as Lipschitz regularization,
was further investigated by \cite{wendlinger_comparative_2024} and led to some limitations of this quantum advantage claim.

\subsection{Physical Fault Injection \& Crosstalk}\label{subsec:physical_fault_injection}
Beyond side-channel information leakage, crosstalk can also be exploited as a noise attack to deliberately disrupt quantum computations.
In this context, \cite{gaur_crosstalk_2025} propose explicit crosstalk-based attacks on trapped-ion quantum computers, where adversarial circuits inject sequences of entangling gates, primarily CNOT operations, to amplify crosstalk effects in a multi-tenant setting. The effectiveness of these attacks is demonstrated using noise-based emulation of trapped-ion devices. In superconducting architectures, noise can be deliberately induced through CNOT drives applied to neighboring qubits \cite{ash-saki_analysis_2020,harper_crosstalk_2024} or by exploiting SWAP paths \cite{lee_swap_2025}.
\subsection{Multi-Stage Attacks and Kill-Chains}
In classical IT security, attacks are rarely executed as isolated actions but rather unfold as structured, multi-stage campaigns. Kill-chain models capture this progression by decomposing adversarial behavior into sequential phases. Prominent examples are the Lockheed Martin Cyber Kill Chain\footnote{\url{https://www.lockheedmartin.com/en-us/capabilities/cyber/cyber-kill-chain.html}} and the MITRE ATT\&CK
framework\footnote{\url{https://attack.mitre.org/}}, which provide a systematic vocabulary for describing attacker objectives,
capabilities, and techniques across different stages. A key insight of kill-chain thinking is that early-stage actions often serve as
prerequisites for later, higher-impact attacks: information gathered during
reconnaissance can enable precise exploitation, while partial access gained early may be leveraged to establish persistence or execute targeted manipulation. As argued in our prior work \cite{debus_entangled_2025}, this perspective is particularly relevant for quantum machine learning systems deployed in cloud-based, multi-tenant environments, where physical-layer effects, control software, and algorithmic components interact in non-trivial ways. Adapting classical kill-chain models to the QML context enables
structured reasoning about how side-channel leakage, crosstalk, noise injection, and
model manipulation can be combined into multi-stage attack scenarios, rather than being analyzed in isolation.

\section{Background}\label{sec:background}

\subsection{Kill-Chain Model for QML}
Drawing on classical kill chain frameworks, we adapt a five-stage model to accommodate the unique architecture and life-cycle of quantum machine learning. These stages, namely Reconnaissance, Initial Access, Model Access / Manipulation, Persistence, and Exfiltration / Impact, cover the necessary steps from an adversary’s early information gathering to the final destructive or exfiltrative actions.

\begin{itemize}
\item\textbf{Reconnaissance stage}
The reconnaissance phase includes general discovery techniques as well as quantum-specific methods for understanding circuit structure and resource usage. In classical IT systems, reconnaissance usually involves scanning for open ports or unpatched services. In contrast, QML contexts present more subtle approaches, such as power or timing side-channels, which can disclose information about circuit structure.

\item\textbf{Initial Access stage}: After the attacker finds a viable entry point, they proceed to the initial access stage, where they establish partial yet significant footholds. This access may involve manipulating scheduling algorithms that assign jobs for securing concurrent execution with victim circuits on a cloud quantum device, or injecting malicious data into a shared dataset.

\item\textbf{Model Access / Manipulation stage}: Once a foothold is established, this stage signifies a shift to actively engaging with the victim's QML model or related circuits. An adversary could alter parameterized gates to reduce accuracy, embed hidden trojans in the transpiled circuit, or introduce extra noise during training. Such manipulation can occur throughout the iterative training loop, particularly in variational circuits, and may remain stealthy if the malicious behavior is triggered only under specific inputs, circuit configurations, or training conditions.

\item\textbf{Persistence stage}: Depending on their overall objective, an attacker may attempt to establish stealthy long-term control mechanisms within the system or the victim model. The most straightforward example of this is Trojan attacks through data poisoning, but other options, such as intentional miscalibration for gate injection, could also be feasible.

\item \textbf{Exfiltration / Impact}: Ultimately, the final stages represent the culmination of the adversary's goals. In some instances, exfiltration involves stealing quantum model parameters or measurement data, thereby replicating proprietary QML logic without compensating for the model's development. In other cases, the main objective is sabotage, such as inducing misclassifications, creating DoS conditions through amplified noise, or exposing private data used in training. This concluding stage highlights that, while multi-stage attacks can occur at any point, the attacker’s primary aim is usually achieved when they either exfiltrate sensitive assets (e.g., trained QML models, confidential data) or cause disruptive effects (e.g., inference failures).
\end{itemize}

For a more detailed discussion of this kill-chain model, we refer the reader to prior work \cite{debus_entangled_2025}.
\subsection{QML Preliminaries}
\subsubsection{Architectures}
In this work we focus on variational classifiers, that is, parameterized quantum circuits for supervised classification tasks, where classical input data are encoded into qubit states, processed through variational layers with trainable parameters, and measured to predict discrete class labels. Upon measurement, the circuit’s outputs feed back into a classical optimizer, such as gradient descent, to refine those same parameters over multiple iterations. To place this concrete choice within the broader design space of quantum machine learning, we adopt the view summarized in Table~\ref{tab:qml-hierarchy}.

\begin{table}[htbp]
  \centering
    \caption{Hierarchy of QML Model Components.}
  \label{tab:qml-hierarchy}
  \begin{tabular}{|l|l|}
    \hline
    \textbf{Category} & \textbf{Illustrative subcategories} \\
    \hline
    QML NISQ Architecture Types &
      \begin{tabular}[c]{@{}l@{}}
        Variational Classifier \\
        Autoencoder \\
        Kernel Methods \\
        Tensor Network Ansatz
      \end{tabular} \\
    \hline
    Architectural Variations &
      \begin{tabular}[c]{@{}l@{}}
        Sequential \\
        Re uploading with Separate Layers \\
        Re uploading with Integrated Layers
      \end{tabular} \\
    \hline
    Data Encoding Strategies &
      \begin{tabular}[c]{@{}l@{}}
        Angle Encoding \\
        Amplitude Encoding \\
        Basis Encoding
      \end{tabular} \\
    \hline
    Layer Types &
      \begin{tabular}[c]{@{}l@{}}
        Basic Entangling Layers \\
        Strongly Entangling Layers \\
        IQP Ansatz
      \end{tabular} \\
    \hline
    Entangling Gate Details &
      \begin{tabular}[c]{@{}l@{}}
        Gate Types \\
        Entanglement Patterns \\
        (Cyclic, Range based)
      \end{tabular} \\
    \hline
    Rotations &
      \begin{tabular}[c]{@{}l@{}}
        Rotation Axis / Pauli Generator \\
        Rotation Angles (and Precision)
      \end{tabular} \\
    \hline
  \end{tabular}
\end{table}

At the top level, we distinguish between different QML NISQ architecture types (variational classifiers, autoencoders, kernel based models, and tensor network inspired ansatzes), which define the overall learning paradigm and circuit structure. Architectural variations specify how data encoding and trainable layers are arranged (purely sequential or with different forms of data re uploading), while data encoding strategies determine how classical features are mapped to quantum states, with direct consequences for circuit depth and width. Layer types control the expressive power and entanglement structure of the repeated building blocks, from basic entangling layers to strongly entangling ones and Instantaneous Quantum Polynomial (IQP) style ansatzes. Finally, entangling gate details (specific two-qubit gate choices and entanglement patterns) together with the choice of rotation generators and angles specify the concrete low level implementation. This illustrates that for a fixed learning task there exists a very large number of possible designs.

\subsubsection{Data Encoding \& Reuploading}

The performance of a QML model on classical data is
heavily dependent on the chosen feature map \cite{ranga_quantum_2024}. This choice represents a trade-off between qubit resources, circuit
depth, and the amount of information encoded. Some techniques for mapping classical input data onto quantum states, affecting circuit depth, width, and learning ability \cite{upadhyay_quantum_2025} are:
\begin{itemize}
    \item \textbf{Angle Encoding:} Features are translated into rotation angles for
    single-qubit gates. This approach is simple and hardware-compatible.
    
    \item \textbf{Amplitude Encoding:} Data are encoded into the amplitudes of a
    quantum state. While compact in terms of required qubits, this method requires complex state preparation.
    
    \item \textbf{Basis Encoding:} Features are mapped to distinct computational basis states, making this encoding suitable for discrete data.
\end{itemize}
The data encoding scheme used in this work is the so-called re-upload encoding introduced by Perez-Salinas et~al. \cite{perez-salinas_data_2020}, where the features of the input vector are grouped into tuples and
sequentially inserted as rotational angles of (general) rotational
gates. The details are covered in Section \ref{subsec:victim_qnn_design}.

\subsection{Power Traces}\label{subsec:power_traces}
Quantum computers implement gate operations via control pulses. For example, superconducting devices use microwave pulses \cite{erata_quantum_2024}, while
AQT trapped-ion systems employ laser pulses. Since these control pulses are generated by classical hardware, they can be intercepted, allowing an attacker to deduce critical circuit details. A central property of quantum hardware is its set of native (or basis) gates, which varies across manufacturers and devices. Before execution, high-level quantum circuits are decomposed into these basis gates. This process, analogous to classical compilation, involves transpiling the circuit at the gate level by breaking down non-native gates into native ones, optimizing the circuit, mapping logical qubits to physical qubits, and finally scheduling the resulting pulses. The final pulse-level circuit description specifies all necessary pulses, including their envelopes, frequencies, phases, and precise timing to ensure the circuit operates correctly on the target quantum device. The attacker aims to uncover quantum circuit details from captured power traces.
\subsection{Crosstalk Mechanisms}\label{subsec:crosstalk_mechanisms}
Quantum crosstalk refers to unintended interactions between
qubits in a quantum processor. These interactions
allow operations on one set of qubits to influence
nearby qubits, thereby reducing fidelity and introducing errors
that degrade computational performance \cite{murali_software_2020}.
\subsubsection{Ion-Trap Crosstalk}
In trapped ion devices quantum gates are implemented using tightly focused laser beams that address individual ions. Each beam generates a localized electric field profile centered on its target ion but with a finite spillover onto neighboring sites due to the diffraction limit. This spatial profile implies that single qubit gates introduce nearest neighbor crosstalk because their beams produce the largest off target fields. In contrast, two qubit gates realized via frequency-detuned laser interactions exhibit strongly suppressed crosstalk since the detuning reduces unwanted excitation of non-addressed ions.
\subsubsection{Superconducting Crosstalk}
In superconducting transmon processors, crosstalk arises from unintended interactions induced by microwave control pulses in the presence of frequency crowding and residual couplings between neighboring qubits. In fixed-frequency architectures, collisions
between qubit transition frequencies and anharmonic transitions can lead to undesired driving of individual qubits or induce spurious interactions during single- and two-qubit
gate operations, thereby degrading computational fidelity \cite{ketterer_characterizing_2023}.
In particular, two-qubit gates constitute a major
source of crosstalk in superconducting devices, in contrast to AQT trapped-ion systems, where
crosstalk is predominantly driven by imperfect single-qubit addressing.

\subsection{Adversarial Examples}\label{subsec:adversarial_examples}
Adversarial examples in (quantum) machine learning are small, often imperceptible perturbations $\delta$, typically constrained by $\|\delta\|_\infty \le \varepsilon$, that are crafted to induce misclassification by maximizing the model’s loss:
\[
\delta^* = \arg\max_{\|\delta\|_\infty \le \varepsilon}
\mathcal{L}\bigl(f(x+\delta;\theta),y\bigr).
\]
Here, $\mathcal{L}(f(x;\theta),y)$ denotes the loss function measuring the discrepancy between the model’s prediction $f(x;\theta)$ and the true label $y$, and $x$ is the original input sample. In practice, this optimization is often approximated using Projected Gradient Descent (PGD), which iteratively increases the loss with respect to the input by taking gradient-ascent steps and then projects the perturbed sample back onto the $\ell_\infty$ ball of radius~$\varepsilon$ centered at the original input. Formally, the PGD update takes the form
\[
x^{(t+1)} = \mathrm{Proj}_{\|\cdot\|_\infty \le \epsilon} \left[ x^{(t)} + \alpha \cdot \mathrm{sign}\bigl(\nabla_{x} \mathcal{L}(f(x^{(t)}), y)\bigr) \right],
\]
where $\alpha$ is the step size. Both classical convolutional networks and variational quantum circuits (employing re-upload or amplitude encodings) have been shown to succumb to such attacks, revealing shared vulnerabilities between QML and classical models \cite{wendlinger_comparative_2024}.

\subsection{Threat Model}
One aspect that is often neglected in current literature on
adversarial QML, but is an established process in classic IT
security, is threat modeling. A threat model that explains typical
threat scenarios and assesses the potential impact according to
the attacker’s capabilities and resources, should be part of any
analysis of security aspects of QML.

\subsubsection{Attacker Capabilities}
Since we perform different experiments, the attacker’s capabilities must be
considered individually for each attack scenario. Table~\ref{tab:attacker_capabilities}
summarizes the required access levels and assumptions.

\begin{table}[htbp]
\centering
\caption{Attacker capabilities and assumptions for the evaluated attack scenarios.}
\label{tab:attacker_capabilities}
\begin{tabular}{p{0.32\linewidth} p{0.64\linewidth}}
\hline
\textbf{Attack Vector} & \textbf{Required Access and Assumptions} \\
\hline
Power Trace Attack &
Requires access to fine-grained analogue power information at the
control-electronics layer of a device. Such access is not
available to ordinary cloud users but represents a realistic threat model
for privileged insiders, on-premise deployments, or future platforms
exposing pulse-level diagnostics. \\
\hline
Adversarial Attacks via Data Poisoning &
Assumes white-box access to model gradients and the ability to repeatedly
query the trained QNN during iterative PGD optimisation. No structural or
hardware-level information is required. The primary computational cost
arises from repeated forward and backward passes. \\
\hline
Adversarial Examples via Crosstalk &
Requires white-box gradient access and repeated query capability, in addition to a multi-tenant execution scenario that enables concurrent circuit execution
and the exploitation of crosstalk effects during the execution of the victim circuit. Furthermore, explicit timing information about the victim circuit is required, as the attacker’s gates must be executed at specific positions during the circuit execution. \\
\hline
\end{tabular}
\end{table}



\subsubsection{Attack Goals}

The goal of the \textbf{power trace attack} is to extract as much information as possible about the victim circuit, such as the number of involved qubits, the executed gate sequence, or the trained parameters of the victim model. The goal of \textbf{adversarial attacks} is to deliberately manipulate the input of a model with minimal perturbations such that the model produces incorrect predictions or misclassifications, without modifying the model parameters or the underlying hardware. The goal of \textbf{adversarial examples via crosstalk} is to physically induce adversarial perturbations through crosstalk, such that concurrent attacker-controlled gate executions distort the effective input seen by the victim model and approximate the effect of analytically computed adversarial examples, without modifying the input data.
\section{Methods}\label{sec:methods}
\subsection{Overall Experimental Pipeline}
\label{subsec:overall_pipeline}

Our experimental methodology follows a deliberately chained, multi-stage attack pipeline that mirrors an adversarial workflow against a deployed QNN. The central objective is to demonstrate how information obtained during an initial reconnaissance phase can be exploited to enable a targeted manipulation of a victim QNN via hardware-induced crosstalk. In the first stage, the attacker performs \emph{side-channel reconnaissance} to infer structural properties of the victim QNN.  In this work, we demonstrate reconnaissance through a power-trace–based analysis. By comparing the observed signatures of the victim QNN against a library of benchmark circuits, the attacker is able to infer key architectural parameters such as the number of qubits, circuit depth, and entangling structure. Crucially, this stage yields an approximate temporal map of the victim circuit execution, allowing the attacker to identify when specific data-encoding rotations are executed. Building on this knowledge, the second stage focuses on the construction of adversarial examples. We first generate conventional adversarial examples using gradient-based methods on the victim model that matches the inferred architecture of the victim QNN.

\newpage However, physical constraints of the hardware significantly restrict how such perturbations can be realized in practice. In particular, on the considered AQT trapped-ion platform, crosstalk effects can be reliably induced only through single-qubit rotations. Since virtual $R_Z$ gates do not induce any physical disturbance, as a consequence, only input features encoded via $R_Y$ rotations can be perturbed by crosstalk. We therefore refine the adversarial generation process by explicitly constraining perturbations to the subspace corresponding to $R_Y$-encoded inputs, resulting in \emph{adversarial examples via crosstalk} that respect the hardware constraints of the device. In the final stage, the attack is executed on real quantum hardware. Using the timing information obtained during reconnaissance, the attacker schedules carefully chosen disturbance operations on neighboring qubits at specific points in time during the execution of the victim QNN. These operations induce controlled crosstalk, with the goal to approximate the previously constructed adversarial perturbations, thereby manipulating the effective input seen by the victim model. 

\subsection{Victim QNN Design}\label{subsec:victim_qnn_design}

\begin{figure}[htbp]
    \centering
    \includegraphics[
        width=0.9\linewidth,
        alt={Dataset overview showing four image classes used for QNN classification, corresponding to the symbols plus, minus, left turnstile and right turnstile}
    ]{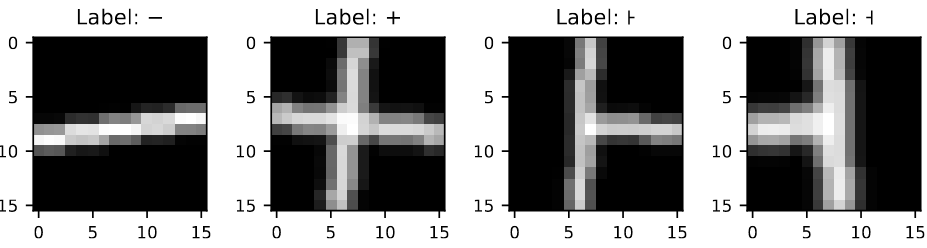}
    \caption{An overview of the used dataset \cite{wendlinger_comparative_2024}. Shown are examples from the individual classes of the dataset.}
    \label{fig:plus_minus_dataset}
\end{figure}

In this section, we introduce the victim QNN, which serves as the target of our attacks detailed in our experimentation plan, and define the benchmark circuits used in the considered attacks. The victim QNN is a modification of the QNN introduced in \cite{wendlinger_comparative_2024}. In the latter implementation, the input features of the QNN are incorporated as rotational angles in the circuit. The QNN’s classification task involves assigning each input image to its respective class. Notably, the dataset depicted in \cref{fig:plus_minus_dataset} comprises four distinct classes: $\{+,-,\vdash, \dashv\}$.

For an input image of dimensions \(16 \times 16 = 256\), the quantum circuit uses 8 qubits and $L=32$ layers. Each layer consists of a block of rotational gates (whose angles are determined by grouped 3-tuples of input features) followed by an entangling layer, which yields a total of 
\[
32 \cdot 8 \cdot 3 = 768
\]
variational parameters. Moreover, each of these 768 rotation parameters has a corresponding input encoding weight, doubling the total number of trainable parameters, leading to a total of \(1536\) trainable parameters in the model.

\begin{figure}[htbp]
    \centering
    \includegraphics[
        width=0.9\linewidth,
        alt={Schematic quantum circuit diagram of a re-upload encoding architecture showing eight qubits arranged in repeated layers of parameterized single-qubit rotations followed by entangling gates, repeated across multiple layers}
    ]{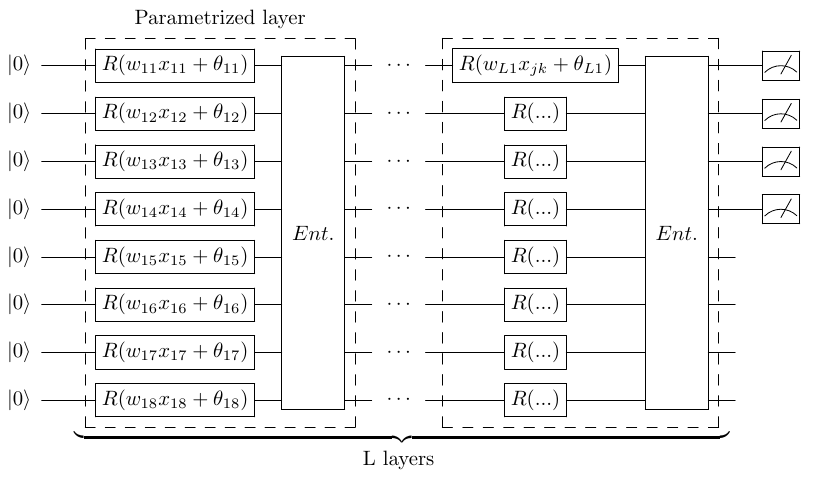}
    \caption{A conceptual overview of the re-upload encoding architecture from reference \cite{wendlinger_comparative_2024}.}
    \label{fig:re-upload-architecture}
\end{figure}

The data encoding scheme used is the so-called re-upload encoding introduced by Perez-Salinas et~al. \cite{perez-salinas_data_2020}. Given some input 3-tuple \(x\) and weights \(w,\, \theta \in \mathbb{R}^3\), this general rotation can be written as a product of unitaries
\[
R(w x + \theta) = U_{w,\theta}(x) = \prod_{j=1}^{3} e^{-i(w_j x_j + \theta_j) H_j},
\]
where the typical gate decomposition of general rotational gates employs the Hamiltonians
\[
H_1 = \frac{1}{2}\sigma_Z, \quad H_2 = \frac{1}{2}\sigma_Y, \quad H_3 = \frac{1}{2}\sigma_Z.
\]
For re-uploading, this implies that each feature of a \(D = 256\)-dimensional input sample is presented to the model three times.  This choice is motivated by the inherent flexibility of the re-uploading paradigm: it decouples the number of qubits from the input dimension, thereby allowing high-dimensional inputs to be encoded repeatedly onto a fixed number of qubits. As model output, a four-dimensional vector is produced by computing the expectation values of the Pauli $Z$ operators on the first four qubits, each corresponding to one of the four target classes. These expectation values serve as the inputs to the $\arg\max$ function, which is used to select the predicted class. The parameter \(r\) defines the entangling structure of the QNN. If \(r\) is not explicitly set (i.e., when \(r\) is specified as \texttt{None}), a default value is computed internally. Specifically, for the \(l\)th layer in a circuit with \(M\) wires, the value of \(r\) is determined as $r = l \mod M.$
Once this value is established, the entangling part of each layer is constructed by applying CNOT gates. For every qubit \(q\), a CNOT gate is applied with qubit \(q\) acting as the control and qubit \((q + r) \mod M\) as the target. In situations where the modulo operation yields zero, \(r\) is adjusted and set to \(1\) to ensure a proper entanglement pattern.

We reduce the complexity of the model by applying Principal Component Analysis (PCA) to the \(16 \times 16\) pixel images, projecting them onto the top 16 principal components and reducing the number of qubits from eight to four. The re-uploading scheme is retained, and the optimized model is implemented with four strongly entangling layers (with \(r\) set to the default value), effectively corresponding to three re-uploads. Through this approach, the number of variational parameters drops from  $32 \times 8 \times 3 \times 2 \;=\; 1536$
to $4 \times 4 \times 3 \times 2 \;=\; 96.$

Similarly, as in the previous model, the output is derived by computing the expectation values of the individual Pauli $Z$ operators for each of the four output qubits. These correspond directly to the four target classes.

\newpage For each input, the quantum circuit returns a four-dimensional vector of $Z$ expectation values, one per output qubit. Classification is then performed by applying the $\arg\max$ function to this vector, selecting the index with the highest value as the predicted class.

We trained the model using state vector simulation to optimize the rotation parameters. Across a total of 1000 training samples and 200 test samples, the model achieved a training accuracy of $0.885$ and a test accuracy of $0.875$. These results reflect an upper bound of accuracy that can be expected using real hardware.

\begin{figure}[h]
    \centering
    \includegraphics[
        width=0.8\linewidth,
        alt={Bar chart showing, for each of the four class labels, the average highest and second highest Pauli Z expectation values predicted by the QNN, illustrating class separation through a clear gap between the maximum and second maximum values}
    ]{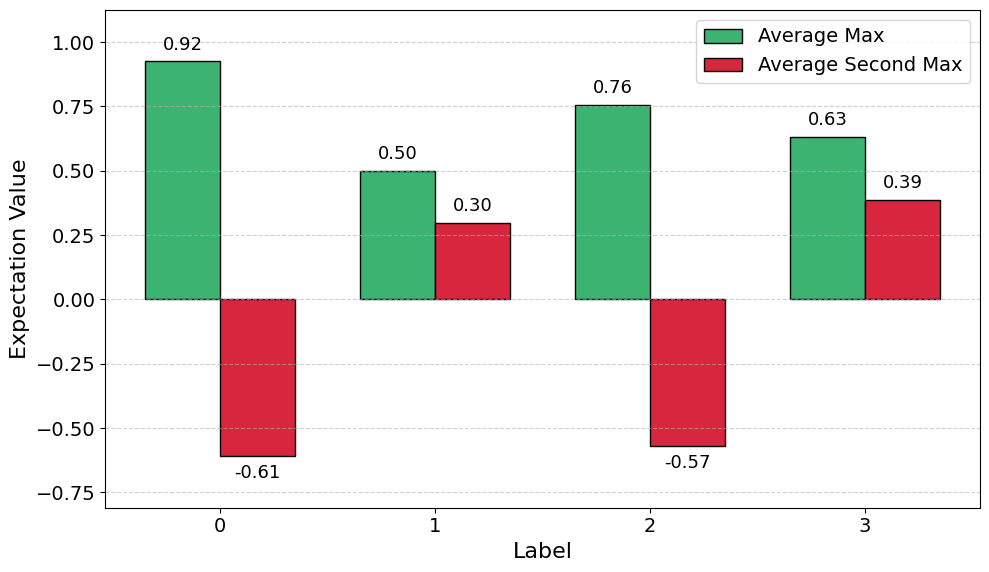}
    \caption{Average highest and second highest Pauli-\(Z\) expectation value (per true label) on the training set.}
    \label{fig:expectation_barplot}
\end{figure}

To better understand the performance of the model for individual classes, we analyze the average highest and second highest expectation values predicted by the QNN for each label (see Fig.~\ref{fig:expectation_barplot}). Here, labels 0 and 2 correspond to the symbols "$-$" and "$\vdash$", while labels 1 and 3 represent "$+$" and "$\dashv$" respectively.

For samples with true labels 0 and 2, the model’s average maximum expectation value is noticeably higher than for labels 1 and 3. At the same time, the average second highest expectation value for classes 1 and 3 is substantially higher than for 0 and 2. This result suggests that the quantum model is, on average, much more confident in its predictions for classes 0 and 2, whereas the model is often less decisive and more ambiguous for classes 1 and 3.

\subsection{Benchmark Circuits}
The main goal of all presented stage one attacks is to infer the structure of a victim QNN by comparing its side-channel signature with those generated by a set of benchmark candidate circuits. For these circuits, no specific input data is provided. Instead, the rotational parameters of each candidate circuit are randomized uniformly between \(0\) and \(2\pi\). In experiments where multiple samples are collected, the rotation angles of the benchmark circuits are re-randomized for each run of the victim QNN, ensuring that a robust and statistically representative side-channel signature is obtained for every candidate.  By comparing the victim's measured side-channel signature with the signatures of the candidate circuits, the attack can identify the most likely underlying circuit structure. When constructing the benchmark circuits, note that the \textbf{constraints}, namely qubit count and circuit depth, are dictated by the underlying quantum computing device.
\begin{itemize}
    \item The number of qubits available on the device determines the maximum number of qubits $q$ that can be used in the circuit. 
    \item The coherence time of the quantum device influences the depth of the circuit. The lower the coherence time, the less deep the circuit should be, as error probabilities increase with circuit depth.
\end{itemize}

These constraints need to be considered when designing the benchmark candidates to ensure that the chosen parameters are feasible for the specific quantum computing architecture being used. We decided to employ the re-uploading scheme exclusively. To evaluate the side-channel signatures, we systematically vary three key parameters, resulting in a total of 24 distinct benchmark configurations:
\begin{itemize}
    \item \textbf{Number of Qubits:} Configurations will be built using either 4 or 8 qubits.
    \item \textbf{Number of Layers:} Circuits will be implemented with 2, 4, 6, and 8 layers.
    \item \textbf{Entanglement Parameter \(r\):} The parameter r will determine the target qubit in the entangling CNOT gates. We will vary \(r\) by selecting explicit values of 1 and 2, as well as by using the default setting. 
\end{itemize}

\subsection{Simulation of AQT Power Traces}
Since real power traces from hardware are not available, we simulate the power traces as follows. To model the power trace on an AQT device, we assign to each qubit \(i\) and qubit pair \((i,j)\) an amplitude coefficient derived from the raw corrections $\{a_i\}_{i=0}^{N-1} = \{1, -1, 2, 3, 0, -2, -3, 0, -4, 4\},$ 
for \(N=10\) qubits.  We then define the single and two qubit scaling factors
$\alpha_i = 1 + \frac{a_i}{100}$ and  $\beta_{ij} = 10 + \frac{a_i + a_j}{10}$, such that \(\alpha_i\) modulates pulses on qubit \(i\) and \(\beta_{ij}\) modulates cross-resonance pulses between qubits \(i\) and \(j\). Each basis gate is represented by a single rectangular pulse of fixed duration followed by a \(2\,\mu\mathrm{s}\) idle, with amplitude determined by the appropriate coefficient:
\begin{itemize}
  \item \(R_Z\) (virtual \(R_Z\)): no physical pulse (\(P=0\)).
  \item \(\mathrm{R}(\theta,\phi)\) (general rotation): one \(10\,\mu\mathrm{s}\) pulse of amplitude \(\alpha_i\,\theta/\pi\), then a \(2\,\mu\mathrm{s}\) idle.
  \item \(\mathrm{R_{XX}}(\theta)\) (XX interaction): one \(200\,\mu\mathrm{s}\) pulse of amplitude \(\beta_{ij}\,\theta/\pi\), then a \(2\,\mu\mathrm{s}\) idle.
\end{itemize}

Given a circuit, we construct the time-ordered trace \(\{(t_k,P_k)\}_{k=0}^K\) by first initializing $t_0 = 0$,  $t_1 = 2\,\mu\mathrm{s}$,  and $P_0 = P_1 = 0$.

For each subsequent gate, we append four entries: two identical amplitude values to represent the start and end of the rectangular pulse, followed by two zeros for the fixed \(2\,\mu\mathrm{s}\) idle. 

\newpage Concretely, since the pulse duration \(\Delta_{\mathrm{pulse}}\) is \(10\,\mu\mathrm{s}\) for single-qubit gates and \(200\,\mu\mathrm{s}\) for two-qubit gates, then
\[
\begin{aligned}
P_k = P_{k+1}
&=
\begin{cases}
\alpha_i\,\dfrac{\theta}{\pi}, & \text{for an \(\mathrm{R}(\theta,\phi)\) on qubit \(i\)},\\[4pt]
\beta_{ij}\,\dfrac{\theta}{\pi}, & \text{for an \(R_{XX}(\theta)\) on \((i,j)\)},
\end{cases}\\
t_{k+1}-t_k &= \Delta_{\mathrm{pulse}}.
\end{aligned}
\]

and
\[
P_{k+2}=P_{k+3}=0,
\quad 
t_{k+2}-t_{k+1} = 0 \,\mu\mathrm{s},
\quad
t_{k+3}-t_{k+2}=2\,\mu\mathrm{s}.
\]
The resulting step function
\[
P(t)=P_k,\quad t_k\le t<t_{k+1},
\]
captures each gate’s amplitude and timing exactly as produced by the AQT power trace simulator.
\subsection{Adversarial Example Generation}\label{subsec:adversarial_example_generation}

In this work, we consider the following threat assumption. Rather than training a surrogate model, we assume white-box access to the victim QNN and directly generate adversarial examples using the trained model in an idealized state vector simulation. This allows us to isolate the effect of hardware-induced noise and implementation-specific deviations on the transferability of adversarial examples, without conflating these effects with surrogate model mismatch. We start with evaluating the robustness of the victim QNN defined in Section \ref{subsec:victim_qnn_design} and systematically construct adversarial examples for every sample in the training set using PGD, as described in \ref{subsec:adversarial_examples}. The procedure iterates over all training samples and searches for the minimum perturbation in $\ell_\infty$-norm that causes the QNN to change its classification relative to the prediction for the unperturbed input. This search is iterative: for each sample, the maximum perturbation $\epsilon$ is increased in increments of $0.1$ up to a maximum of $1.0$, and the smallest $\epsilon$ for which a prediction change occurs is recorded. The evaluation workflow is as follows:
\begin{enumerate}
    \item For each sample $x$ from the set of training data, the model's predicted label is computed.
    \item We initialize the perturbation size $\epsilon=0.1$ and iteratively apply the PGD attack with increasing $\epsilon$ until the perturbed sample $x_{\mathrm{adv}} = x + \delta$ is differently classified or $\epsilon$ reaches $1.0$.
    \item Once successful, the corresponding $\epsilon$ and resulting adversarial sample are stored.
\end{enumerate}

The hyperparameters used for PGD-based adversarial example generation are summarized in Table~\ref{tab:pgd_hyperparameters}.

\begin{table}[htbp]
    \centering
    \caption{Hyperparameters used for PGD-based adversarial example generation.}
    \label{tab:pgd_hyperparameters}
    \begin{tabular}{ll}
        \toprule
        \textbf{Hyperparameter} & \textbf{Value} \\
        \midrule
        Number of iterations $T$ & 100 \\
        Step size $\alpha$ & 0.01 \\
        Initial perturbation budget $\epsilon$ & 0.1 \\
        Perturbation budget increment & 0.1 \\
        Maximum perturbation budget $\epsilon_{\max}$ & 1.0 \\
        Norm constraint & $\ell_\infty$ \\
        \bottomrule
    \end{tabular}
\end{table}


\subsection{Constraint Adversarial Example Generation}\label{subsec:constraint_adversarial_example_generation}
We aim to induce adversarial examples through crosstalk by applying $R_Y$ rotation gates on qubits adjacent to the target. On AQT hardware crosstalk is induced by single-qubit rotation gates. $R_Z$ gates are implemented virtually on the AQT Ibex device, so adding extra $R_Z$ rotations does not produce crosstalk. Consequently we restrict perturbations to the input components that are encoded in the $R_Y$ gates. 
\subsubsection{Implementation}
Let \(x^{(0)} \in \mathbb{R}^{16}\) denote the original input vector. We form the
expanded input \(x \in \mathbb{R}^{D}\) used by the QNN by concatenating
\(x^{(0)}\) three times,
\[
x = \bigl( x^{(0)}_1,\ldots,x^{(0)}_{16},\
x^{(0)}_1,\ldots,x^{(0)}_{16},\
x^{(0)}_1,\ldots,x^{(0)}_{16} \bigr) \in \mathbb{R}^{48},
\]
that is \(D = 3 \cdot 16\). The entries of this expanded vector are assigned to
the encoding gates in the repeating order \((R_Z,R_Y,R_Z)\). In particular, the
first triple consists of \((x^{(0)}_1,x^{(0)}_2,x^{(0)}_3)\), not three copies of
\(x^{(0)}_1\).

We define a binary mask \(m \in \{0,1\}^{D}\) that selects exactly the middle element of each \((R_Z, R_Y, R_Z)\) triple,
\[
m_k =
\begin{cases}
1, & \text{if } k \in \{2,5,8,\dots,47\} \\
0, & \text{otherwise.}
\end{cases}
\]
This mask enforces that only components encoded in \(R_Y\) rotations may be perturbed. Given a differentiable loss \(\mathcal{L}(f(x),y)\) for model \(f\) and label \(y\), a masked PGD step with step size \(\alpha>0\) takes the form
\[
g^{(t)} = m \;\odot\; \mathrm{sign}\!\bigl(\nabla_{x}\mathcal{L}(f(x^{(t)}),y)\bigr), \qquad
\tilde{x}^{(t+1)} \;=\; x^{(t)} \;+\; \alpha\, g^{(t)},
\]
followed by projection onto the \(\ell_{\infty}\) ball of radius \(\epsilon\)
around the unperturbed expanded input \(x\),
\[
x^{(t+1)} \;=\; \mathrm{Proj}_{{\|\cdot\|_\infty \leq \epsilon}}\bigl[\tilde{x}^{(t+1)}\bigr].
\]

We choose the same hyperparameters for the PGD procedure as in Section \ref{subsec:adversarial_example_generation}. For each evaluated training sample, success is declared if the predicted class label changes after the perturbation is applied. In this case, the current $\epsilon$ is stored as the attack threshold for that sample. 

\subsubsection{Determine which $R_Y$ gates need to be applied}

Since our goal is to induce the intended effect in the QNN by exploiting crosstalk, we must determine which rotation gates to apply and where in the circuit to apply them in order to achieve the desired outcome.  In what follows we derive these crosstalk factors exclusively from the AQT Ibex experiment described in Appendix \ref{sec:aqt_crosstalk_experiment}.

Specifically, we place the victim qubits on the ions $[3,5,7,9]$ and use the measurements of the five neighbor ions 2, 4, 6, 8, and 10 recorded in that experiment. For each measured qubit $j\in\{2,4,6,8,10\}$ we estimate an effective $R_Y$ angle from the observed bitstrings. Let $p_1$ denote the proportion of outcomes equal to one at the bit position that corresponds to qubit $j$. 

Consider a single qubit initially prepared in the state $\ket{0}= \begin{pmatrix}
    1 & 0
\end{pmatrix}^T$ . When a rotation about the $Y$-axis by an angle $\theta_{\mathrm{eff},j}$ is applied, the state transforms as  $R_Y(\theta_{\mathrm{eff},j})|0\rangle =
\begin{pmatrix}
\cos\!\big(\tfrac{\theta_{\mathrm{eff},j}}{2}\big) &
\sin\!\big(\tfrac{\theta_{\mathrm{eff},j}}{2}\big)
\end{pmatrix}^T.$

The probability of obtaining the outcome $|1\rangle$ when measuring in the computational basis 
is given by the squared magnitude of the lower amplitude:
\[
p_1 = |\langle 1|R_Y(\theta_{\mathrm{eff},j})|0\rangle|^2
     = \sin^2\!\Big(\tfrac{\theta_{\mathrm{eff},j}}{2}\Big).
\]
Solving this expression for $\theta_{\mathrm{eff},j}$ yields
\[
\sin\!\Big(\tfrac{\theta_{\mathrm{eff},j}}{2}\Big) = \sqrt{p_1}
\quad\Longrightarrow\quad
\theta_{\mathrm{eff},j} = 2\,\arcsin\!\big(\sqrt{p_1}\big).
\]

For a given circuit with per gate angle $\alpha$ applied $r$ times in sequence we define the total applied angle
\[
\theta_{\mathrm{theo}} \;=\; r\,\alpha,
\]
and the crosstalk factor on qubit $j$
\[
f_j \;=\; \frac{\theta_{\mathrm{eff},j}}{\theta_{\mathrm{theo}}}.
\]

Qubits $Q2$ and $Q10$ each have a single adjacent qubit, namely $Q3$ and $Q9$ respectively, whereas $Q4$, $Q6$, and $Q8$ each have two adjacent qubits. Under a simple additive crosstalk model as a working assumption, the net induced rotation on a qubit scales approximately with the number of active neighbors. This explains why the central qubits exhibit larger factors than the edge qubits. If desired, one can normalize the theoretical angle by the number of adjacent attacker qubits, that is use $\theta_{\mathrm{theo}} = r\,\alpha$ for $Q2$ and $Q10$ and $\theta_{\mathrm{theo}} = 2r\,\alpha$ for $Q4$, $Q6$, and $Q8$, which makes crosstalk factors more comparable. The smaller ion spacing toward the chain center qualitatively explains stronger crosstalk but is not included in our normalization. We keep $r$ equal to $10$ for $Q2$ and $Q10$. The factor of two for $Q4$, $Q6$, and $Q8$ comes purely from the neighbor count normalization under the additive model.

The Table \ref{tab:crosstalk_factors_all_probes} below reports these factors for the probe qubits $Q2, Q4, Q6, Q8,$ and $Q10$ across all benchmarks.

\begingroup
\renewcommand{\arraystretch}{1.15}
\begin{table}[htbp]
    \caption{Crosstalk factors $f_j$ for qubits $Q2, Q4, Q6, Q8,$ and $Q10$ across benchmark circuits.}
    \label{tab:crosstalk_factors_all_probes}
    \centering
    \begin{tabular}{rcccccc}
        \toprule
        \textbf{Angle} & \textbf{Total angle} & \textbf{$Q2$} & \textbf{$Q4$} & \textbf{$Q6$} & \textbf{$Q8$} & \textbf{$Q10$} \\
        \midrule
        $\pi/100$ & $\pi/10$   & 0.201 & 0.071 & 0.071 & 0.071 & 0.142 \\
        $\pi/8$   & $10\pi/8$  & 0.016 & 0.013 & 0.011 & 0.013 & 0.016 \\
        $\pi/4$   & $10\pi/4$  & 0.006 & 0.010 & 0.011 & 0.010 & 0.006 \\
        $3\pi/8$  & $30\pi/8$  & 0.009 & 0.008 & 0.011 & 0.008 & 0.008 \\
        $\pi/2$   & $10\pi/2$  & 0.007 & 0.010 & 0.010 & 0.010 & 0.007 \\
        $5\pi/8$  & $50\pi/8$  & 0.008 & 0.009 & 0.011 & 0.011 & 0.009 \\
        $3\pi/4$  & $30\pi/4$  & 0.007 & 0.009 & 0.011 & 0.011 & 0.007 \\
        $7\pi/8$  & $70\pi/8$  & 0.006 & 0.010 & 0.011 & 0.011 & 0.007 \\
        \bottomrule
    \end{tabular}
\end{table}
\endgroup

The measurements indicate that $f_j\approx 0.01$ in the majority of cases. Accordingly, for simplicity we set $f_j=0.01$ for all $j.$ 

Each of the 16 input dimensions is encoded in exactly one $R_Y$ gate in the order
\[
[2,\,5,\,8,\,11,\,14,\,1,\,4,\,7,\,10,\,13,\,16,\,3,\,6,\,9,\,12,\,15].
\]

The mapping from the 16 input dimensions to the $R_Y$ gates is shown in Table
\ref{tab:ry_encoding_order}. The entries are the indices of the original
input vector.

\begin{table}[h!]
    \centering
        \caption{Encoding order of input dimensions for the $R_Y$ gates by layer and qubit.}
    \label{tab:ry_encoding_order}
    \begin{tabular}{ccccc}
        \toprule
        \textbf{Layer} &  \textbf{$Q3$} & \textbf{$Q5$} & \textbf{$Q7$} & \textbf{$Q9$} \\
        \midrule
        1 & 2 & 5 & 8 & 11 \\
        2 & 14 & 1 & 4 & 7 \\
        3 & 10 & 13 & 16 & 3 \\
        4 & 6 & 9 & 12 & 15 \\
        \bottomrule
    \end{tabular}
\end{table}

See \cref{subsec:victim_qnn_design} for details on the re-upload encoding, architecture.

The masked PGD procedure described above yields an adversarial input vector
\(x^{\mathrm{adv}}\) whose nonzero perturbations are restricted to those components that are encoded via \(R_Y\) rotations. In order to realize this adversarial effect on quantum hardware through crosstalk, these abstract input-space perturbations must be translated into concrete physical operations applied to neighboring qubits. Each input dimension \(x_i\) is encoded into exactly one \(R_Y\) gate of the
re-uploading circuit. Denoting by \(w_i\) and \(b_i\) the weight and bias parameters
of the corresponding encoding gate, the effective rotation angle on the target
qubit is given by
\[
\theta_i \;=\; w_i\,x_i + b_i.
\]
For an adversarial input \(x^{\mathrm{adv}}\), the desired change in the effective
rotation angle is therefore
\[
\Delta \theta_i
\;=\;
w_i\,(x^{\mathrm{adv}}_i - x_i).
\]
This quantity represents the angle offset that must be induced on the target
\(R_Y\) gate in order to emulate the adversarial input at the circuit level.

We assume a linear crosstalk model in which a single-qubit \(R_Y\) rotation applied on a neighboring qubit induces a proportional, but attenuated, rotation on the target qubit. For a neighbor qubit \(j\), this relation is modeled as
\[
\theta_{\mathrm{induced},j}
\;=\;
f_j\,\gamma_j,
\]
where \(\gamma_j\) is the applied rotation angle on the neighbor and \(f_j\) is a
hardware-specific crosstalk factor estimated experimentally.

For a target qubit with two adjacent neighbors, the total induced rotation is
assumed to be additive,
\[
\theta_{\mathrm{induced}}
\;=\;
f_{\mathrm{left}}\,\gamma_{\mathrm{left}}
+
f_{\mathrm{right}}\,\gamma_{\mathrm{right}}.
\]

To avoid overfitting to device-specific variations, we adopt a uniform crosstalk
factor \(f_j = f\) for all neighbors. Under this assumption, applying identical
rotations \(\gamma\) on both adjacent qubits yields
\[
\theta_{\mathrm{induced}} \;=\; 2f\,\gamma.
\]

To induce a desired target offset \(\Delta \theta_i\), the required neighbor
rotation angle is therefore
\[
\gamma
\;=\;
\frac{\Delta \theta_i}{2f}.
\]
These rotations are applied symmetrically to all qubits adjacent to the target
qubit hosting the corresponding \(R_Y\) gate.

\paragraph{Amplitude saturation and gate splitting}

Due to the native pulse implementation on the AQT hardware, requested
\(R_Y\) rotations with magnitude larger than \(\pi\) do not yield stronger physical
pulses. Instead, they are decomposed into equivalent equatorial rotations
sandwiched by virtual \(R_Z\) gates, which do not contribute to crosstalk.
Consequently, neighbor rotations are saturated at magnitude \(\pi\),
\[
\gamma_{\mathrm{applied}}
\;=\;
\mathrm{sign}(\gamma)\,\min\!\bigl(|\gamma|,\pi\bigr).
\]

If \(|\gamma| > \pi\), the required disturbance is implemented as a sequence of
multiple \(R_Y(\pm\pi)\) rotations followed by a final residual rotation, ensuring
that the cumulative induced effect approximates the desired \(\Delta \theta_i\).

\section{Results}\label{sec:results}
In line with the methodological structure introduced in Section~\ref{sec:methods}, we present and discuss the results according to the same multi-stage attack pipeline.
\subsection{Stage 1: Side-Channel Reconnaissance}\label{subsec:side_channel_reconnaissance}

Before generating the power trace, the victim QNN circuit is transpiled into the native AQT gates with \verb|optimization_level=0| to ensure that every gate is preserved exactly as written, without any simplification or merging.
Once the circuit is in this basis, we execute the power trace simulator to generate \(\{(t_k,P_k)\}_{k=0}^K\).

\subsubsection{Experimental Results}
We begin by fixing a single training sample and computing its corresponding victim QNN power trace
\[
\bigl(\{t_k^{(v)}\},\,\{P_k^{(v)}\}\bigr).
\] Next, for each benchmark circuit \(j\in\{1,2,\dots,24\}\), we computed the corresponding power trace \(\bigl(\{t_k^{(j)}\},\{P_k^{(j)}\}\bigr)\). We then compare these traces in three steps:
\paragraph{1. Timing Filter}
We first eliminate any benchmarks whose total execution duration differs from the victim’s trace. Concretely, we compute
\[
T^{(j)} = \max(t_k^{(j)}), 
\quad
T^{(v)} = \max(t_k^{(v)}),
\]
and retain only those \(j\) for which \(T^{(j)} = T^{(v)}\). In our case, this reduces the original 24 candidates to just three, namely the benchmark circuits composed of four layers acting on four qubits.
\paragraph{2. Distance Metric}  
Let 
\[
\Delta_{\mathrm{step}} \;=\; 1\,\mu\mathrm{s}.
\]
For each original interval \([\,t_{k-1},\,t_{k})\) with \(k=1,\dots,K\), define
\[
M_k \;=\; \frac{\,t_{k} - t_{\,k-1}\,}{\Delta_{\mathrm{step}}} \in \mathbb{N} ,
\]
and introduce the intermediate times
\[
\tilde{t}_{k,m} \;=\; t_{\,k-1} \;+\; m\,\Delta_{\mathrm{step}}, 
\quad
m = 0,1,2,\dots,\,M_k - 1,
\]
so that \(\tilde{t}_{k,m} < t_{\,k}\) for all \(m\). For the case $k = K$, we append a single final point at the very end of the trace:
\[
\tilde{t}_{\,K,M_K} \;=\; t_{\,K}, 
\qquad 
\text{and assign } \tilde{P}_{\,K,M_K} = P_{\,K}.
\]
Thus, the total number of fine-grid points is
\[
N \;=\; \sum_{k=1}^{K} M_k \;+\; 1,
\]
where the “\(+\,1\)” accounts for the endpoint \((t_K,P_K)\).

We enumerate all points on the fine grid \(\{\tilde{t}_n\}_{n=0}^N\) in lexicographic order on the pairs \((k,m)\). 
With this definition, the forward-filled amplitude at each \(\tilde{t}_n\) is simply
\[
\tilde{P}_n \;=\; P_{\,k}\quad\text{whenever }\tilde{t}_n\in [\,t_{\,k-1},\,t_k)\,, 
\quad
\tilde{P}_{\,N} = P_{\,K}.
\]
Denote the resulting discrete sequences by
\(\{P_n^{(v)}\}_{n=0}^N\) and \(\{P_n^{(j)}\}_{n=0}^N\). We then compute for each candidate \(j\) the Euclidean distance
\begin{equation}\label{eq:euclidean_distance}
  d_j
  = \sqrt{\sum_{n=0}^{N}\bigl(P_n^{(j)} - P_n^{(v)}\bigr)^2},
\end{equation}
which quantifies the total \(L^2\)-deviation between the two power profiles on the same microsecond-resolution grid.

\paragraph{3. Best Match}  
Sorting the distances \(\{d_j\}\) in ascending order yields
\[
\begin{array}{r@{\;}l}\label{arr:l_2_distances}
d_7 &\approx 9.84,\\
d_9 &\approx 119.14,\\
d_8 &\approx 132.14.
\end{array}
\]
Thus, benchmark circuit 7 has the smallest distance and is selected as the best match, confirming that it indeed corresponds to the victim QNN’s structure. Note that the concrete distance values depend on the specific \(R_Y\) rotation
parameters and on the concrete training sample encoded in the QNN. In our
evaluated instances, circuit 7 consistently yields the smallest
\(L^2\)-deviation among benchmark circuits 7, 8, and 9, indicating that it is the
best structural match to the victim QNN. Since the attacker can infer the concrete structure of the victim QNN, they can now leverage this information to place targeted disturbance gates on neighboring qubits at specific points in time during the execution of the victim QNN. The concrete simulated power traces corresponding to the \(L^2\)-deviation displayed in \ref{arr:l_2_distances} are shown in Appendix \ref{sec:simulated_aqt_power_traces}.
\subsection{Stage 2: Adversarial Examples}\label{subsec:adversarial_examples}
\begin{figure}[htbp]
    \centering
    \setlength{\tabcolsep}{2pt}
    \renewcommand{\arraystretch}{0.9}
    \begin{tabular}{ccccc}
        \includegraphics[
            width=0.15\linewidth,
            alt={Grid of side by side visualizations comparing original images and their adversarial counterparts reconstructed from PCA space to the 16 by 16 pixel domain, for multiple perturbation strengths epsilon ranging from 0.1 to 1.0, illustrating progressively stronger pixel level distortions as epsilon increases}
        ]{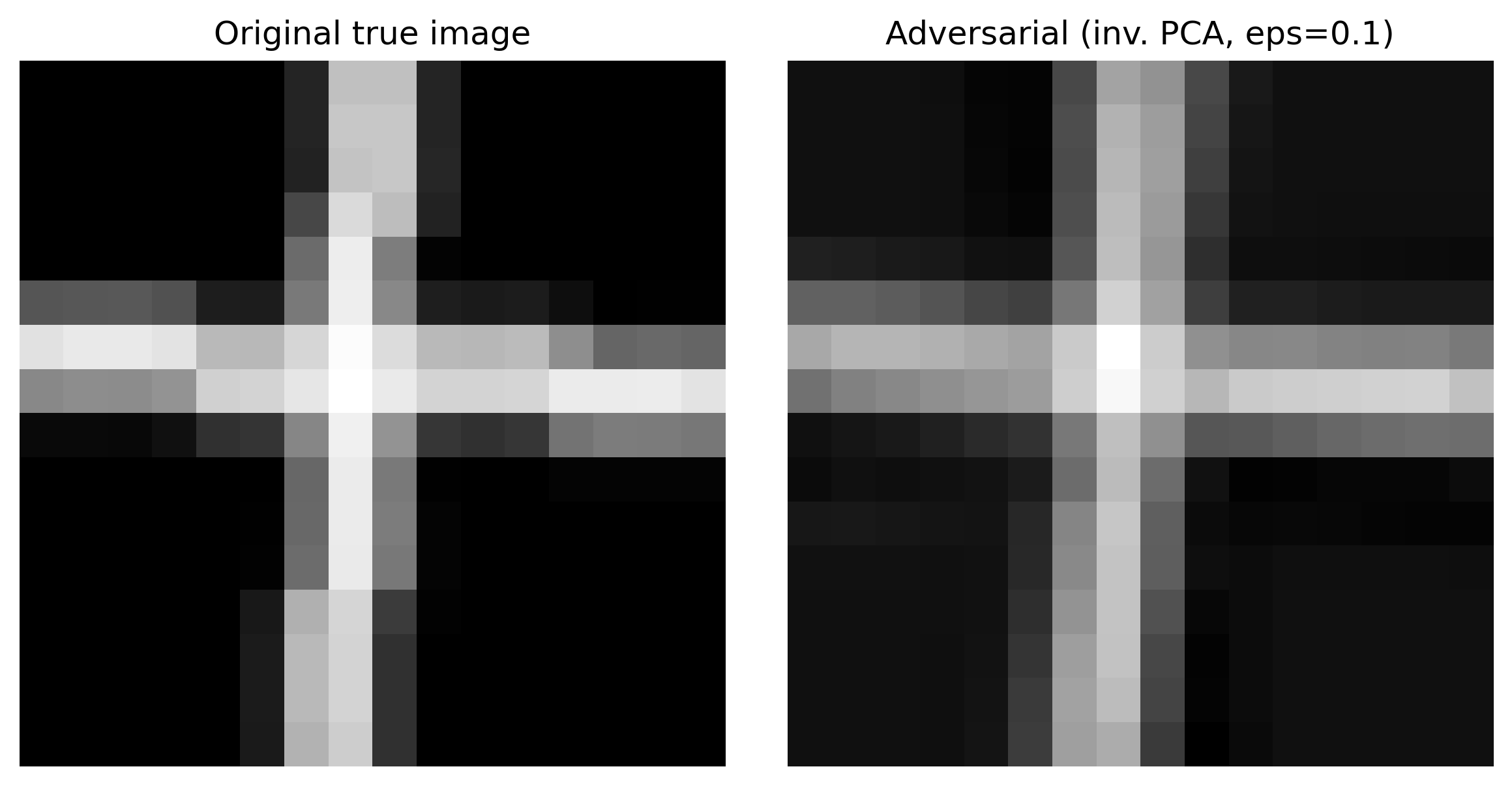} &
        \includegraphics[width=0.15\linewidth]{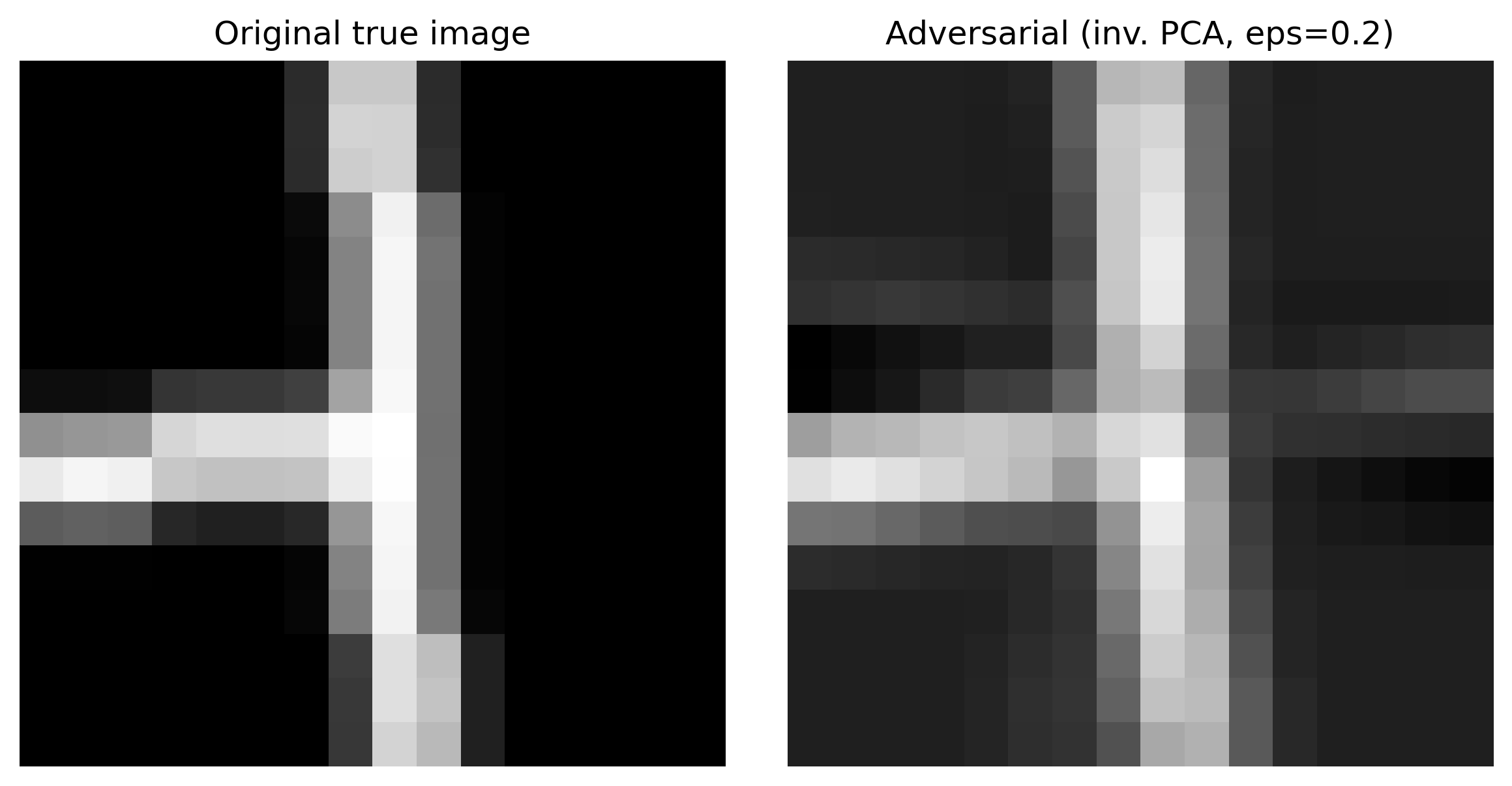} &
        \includegraphics[width=0.15\linewidth]{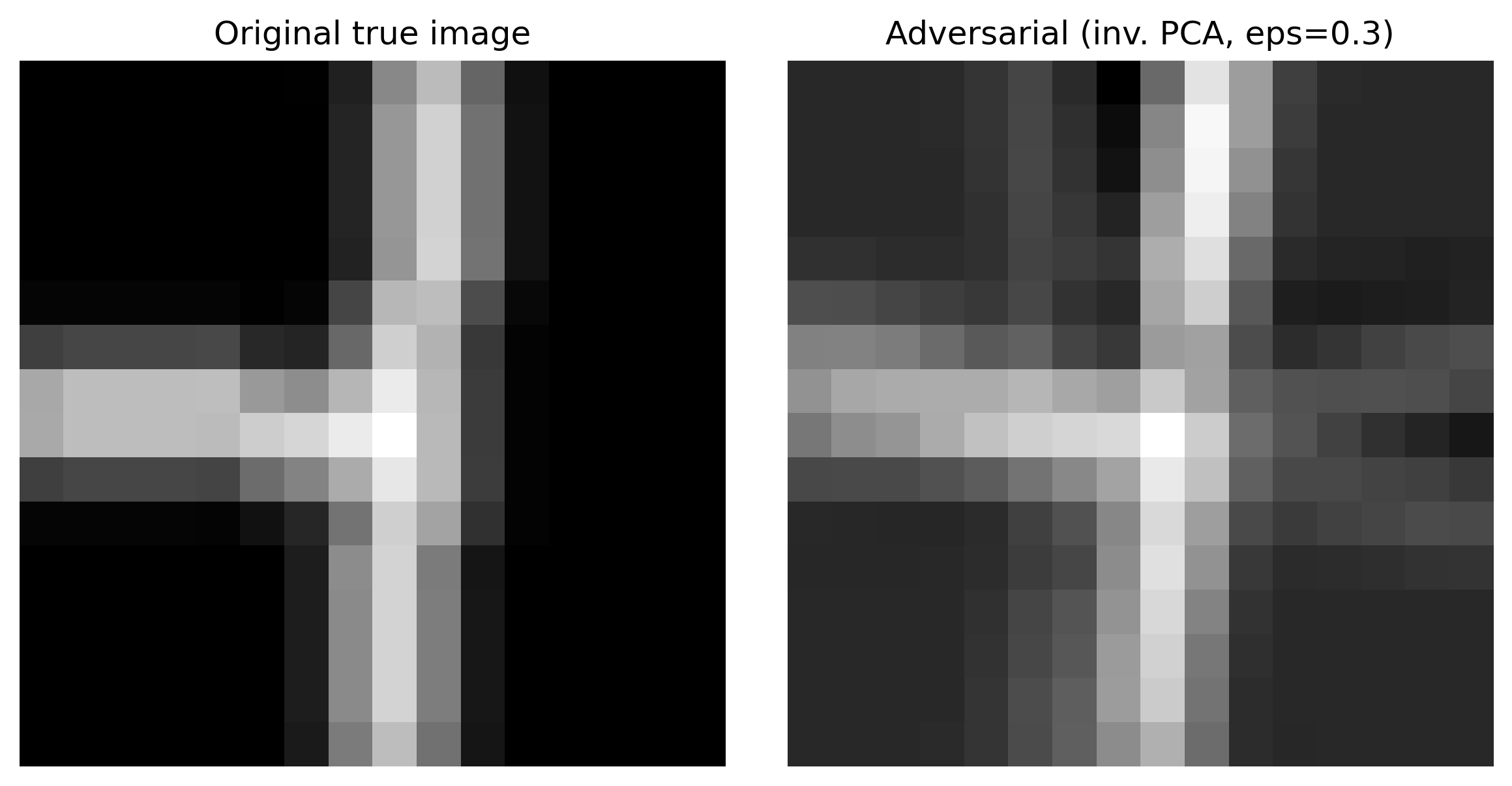} &
        \includegraphics[width=0.15\linewidth]{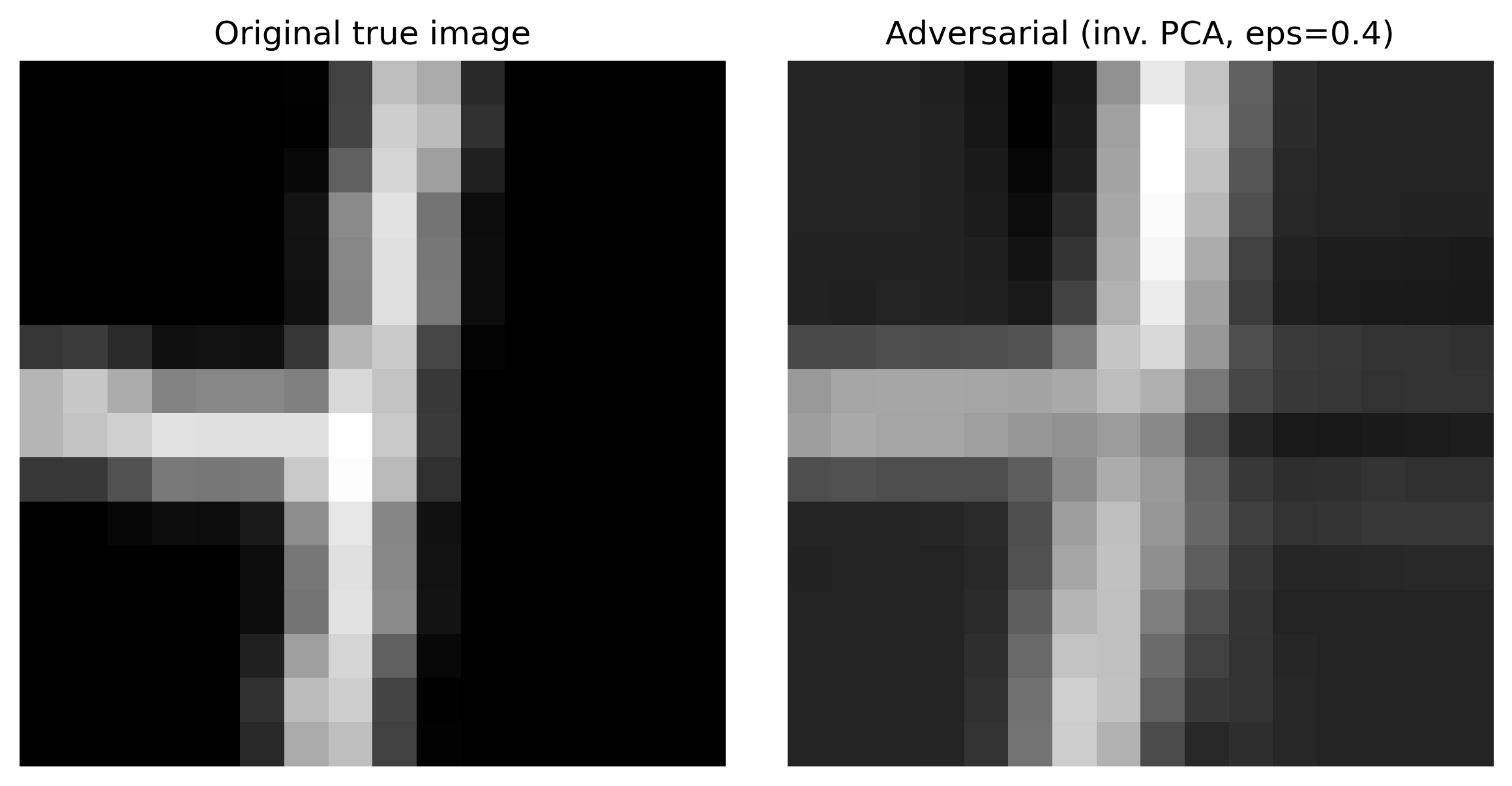} &
        \includegraphics[width=0.15\linewidth]{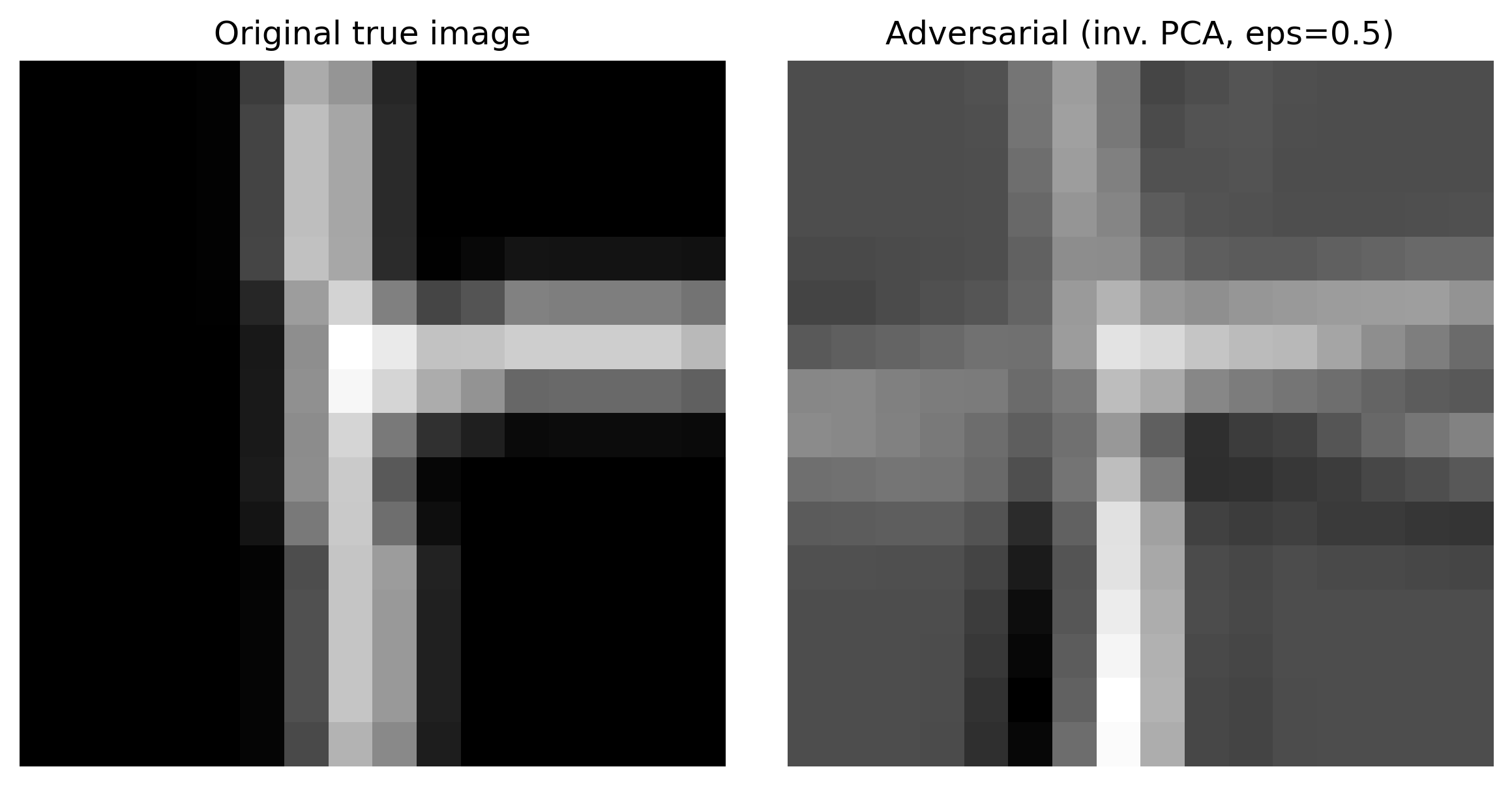} \\
        \small $\epsilon{=}0.1$ &
        \small $\epsilon{=}0.2$ &
        \small $\epsilon{=}0.3$ &
        \small $\epsilon{=}0.4$ &
        \small $\epsilon{=}0.5$ \\
        \includegraphics[width=0.15\linewidth]{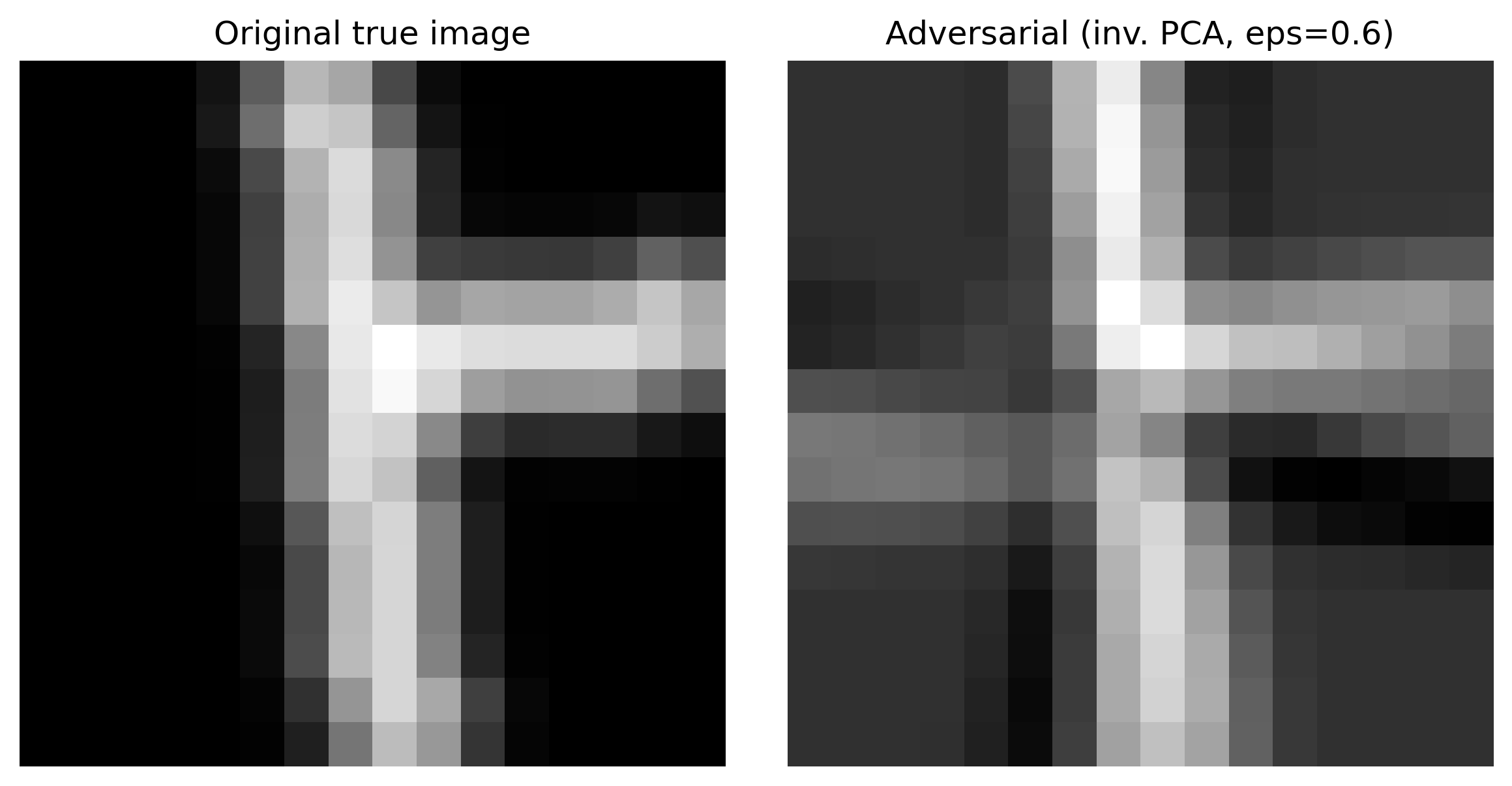} &
        \includegraphics[width=0.15\linewidth]{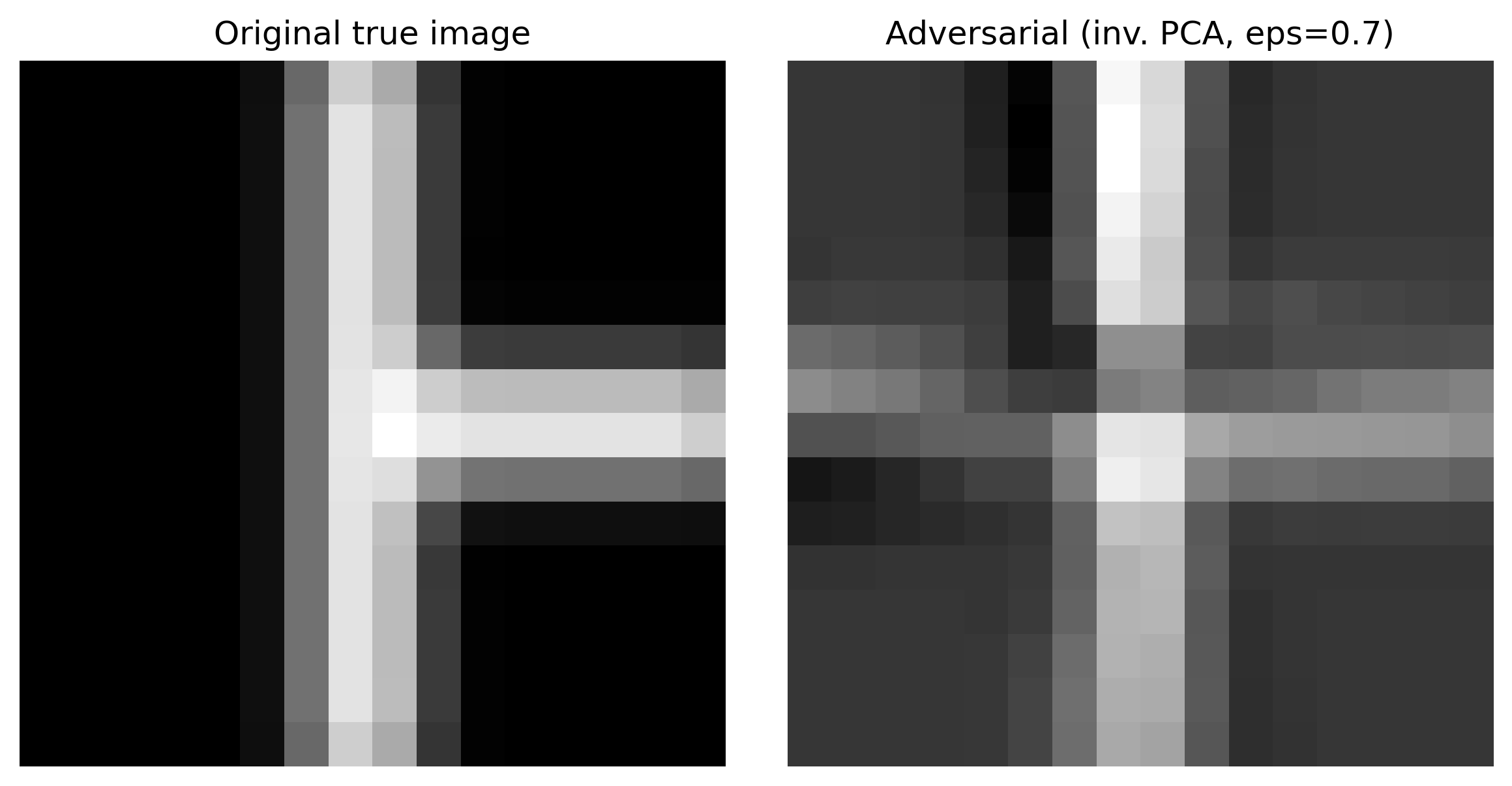} &
        \includegraphics[width=0.15\linewidth]{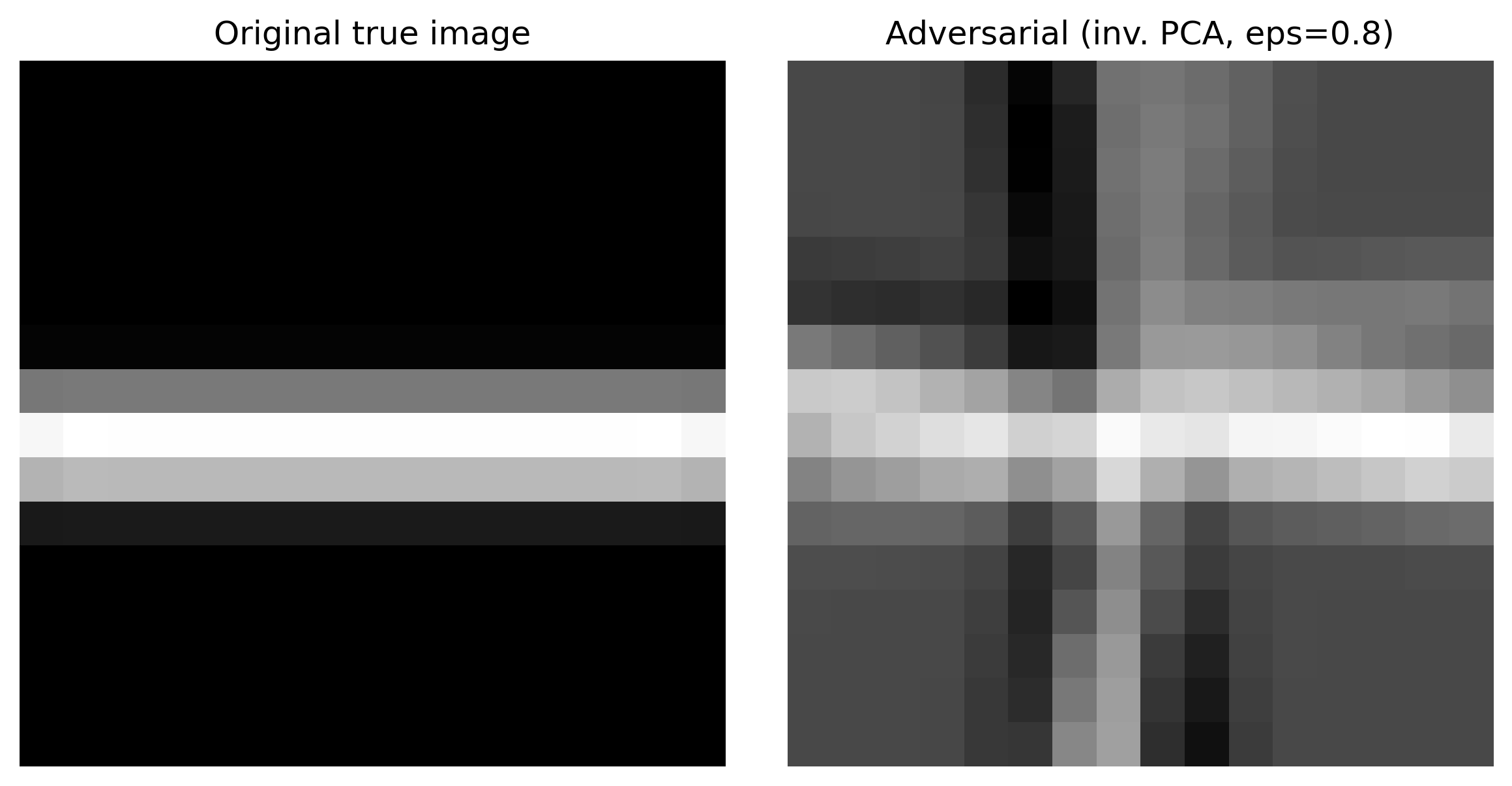} &
        \includegraphics[width=0.15\linewidth]{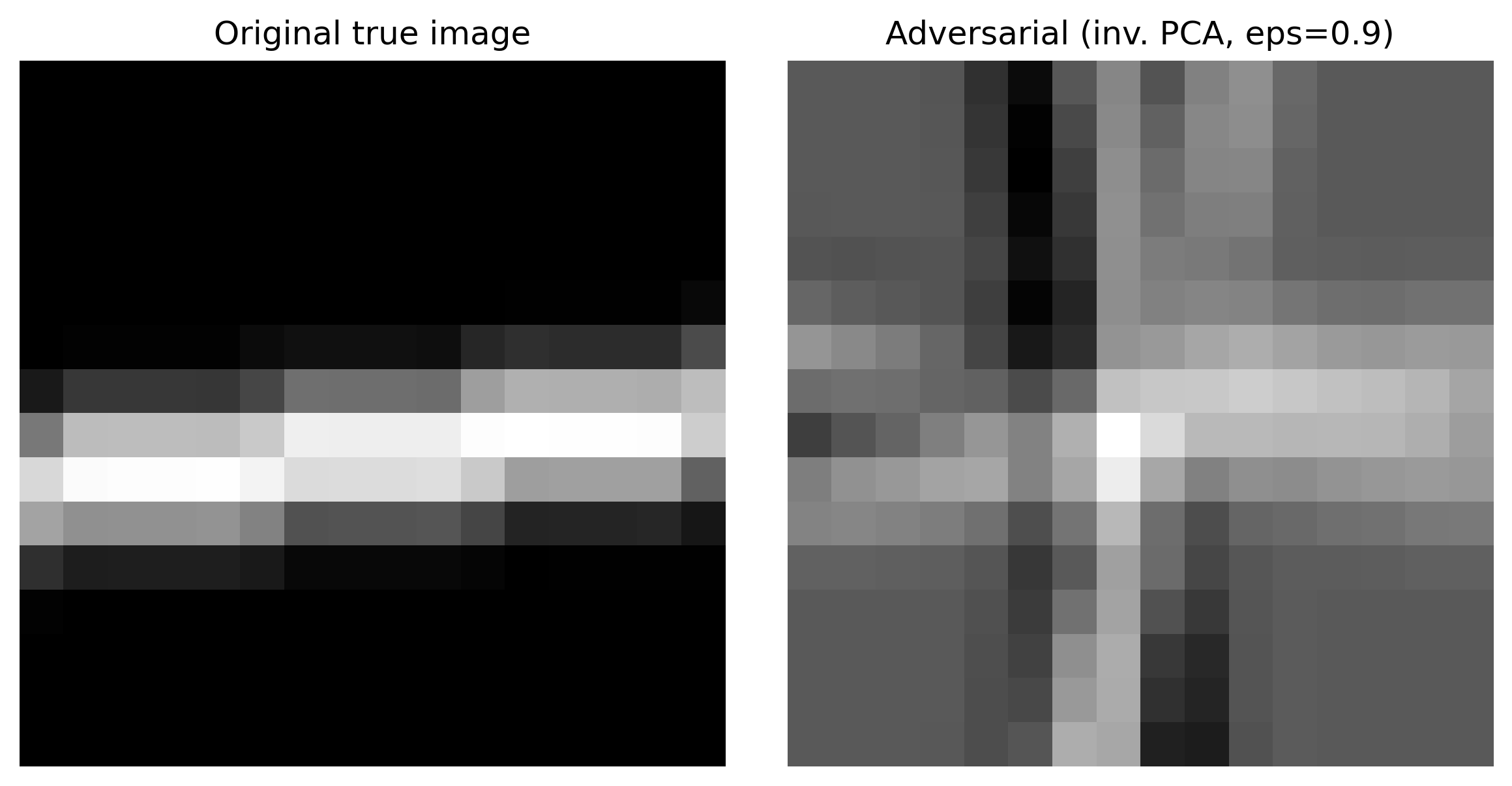} &
        \includegraphics[width=0.15\linewidth]{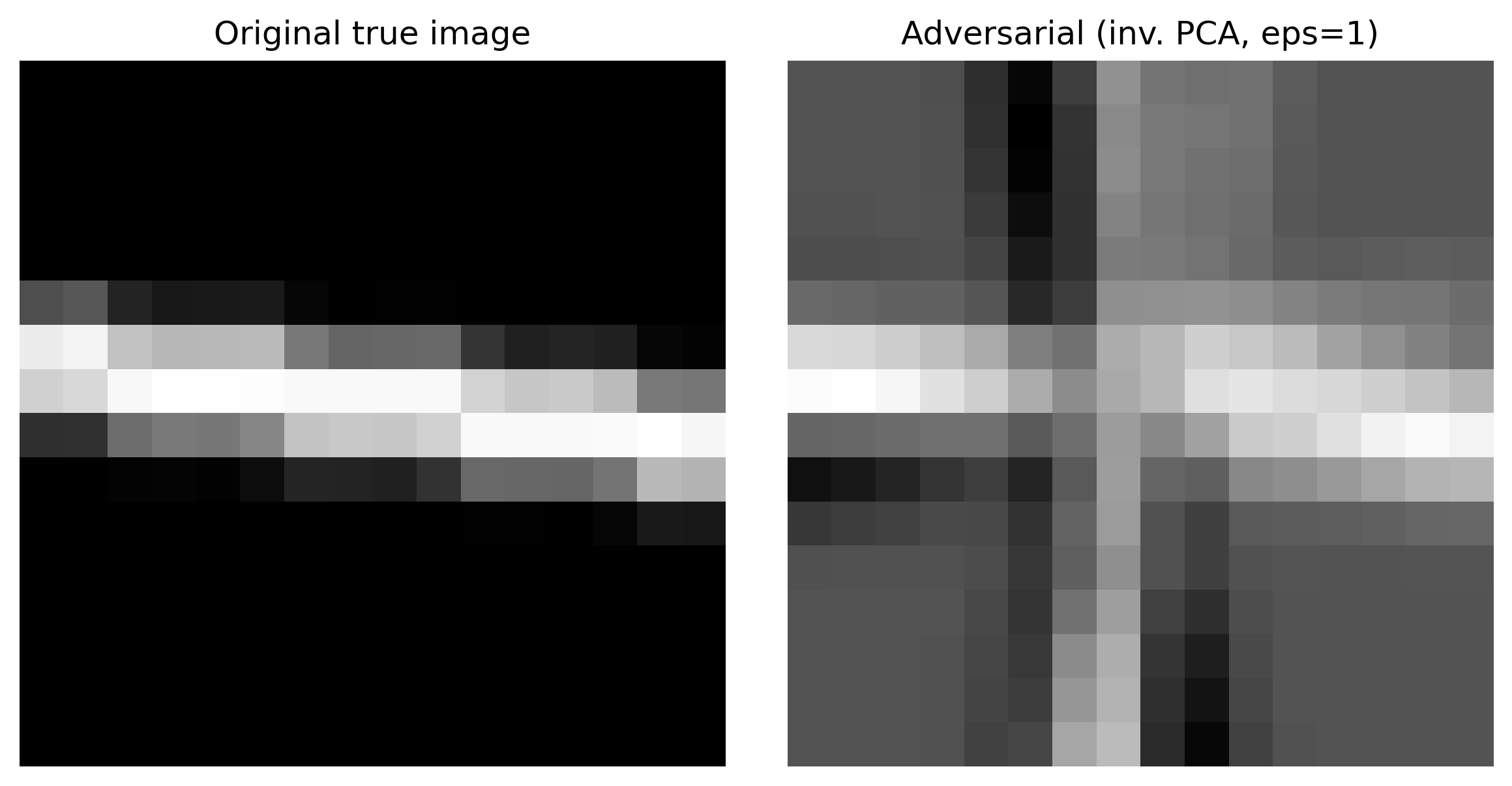} \\
        \small $\epsilon{=}0.6$ &
        \small $\epsilon{=}0.7$ &
        \small $\epsilon{=}0.8$ &
        \small $\epsilon{=}0.9$ &
        \small $\epsilon{=}1.0$ \\
    \end{tabular}
    \caption{Overview of side-by-side visualizations of original (left) and adversarial (right) samples for various attack strengths $\epsilon$. Each panel shows the reconstructed pair from PCA space to the $16\times 16$ pixel domain for a selected training sample.}
    \label{fig:adv_examples_overview}
\end{figure}

As discussed in Section \ref{subsec:adversarial_example_generation} we use the PGD algorithm to generate adversarial examples for the 1000 training samples. To visualize the effect of such adversarial perturbations in the original image space, we performed an inverse PCA transformation of the adversarial examples back into the $16 \times 16$ pixel domain. This enables a  visual comparison between the true image and its adversarially perturbed counterpart. We repeat this reconstruction for several samples at different perturbation magnitudes $\epsilon$. For each case, Figure~\ref{fig:adv_examples_overview} shows a side-by-side visualization of the original image and its corresponding adversarial variant in the pixel domain.

The following Table \ref{tab:epsilon-histogram} presents the distribution of the minimum $\epsilon$ values required to generate successful adversarial examples for the training set using PGD, as described above. For each sample, the smallest $\epsilon \leq 1.0$ that sufficed to trigger misclassification is recorded. The column \textbf{Original correct} indicates how many of these adversarial examples were generated from samples that the QNN originally classified correctly and were subsequently misclassified after adversarial perturbation. The column \textbf{Label flipping} counts cases where the original prediction was incorrect, but the adversarial example caused the prediction to switch to the correct class. Notably, for $160$ out of $1000$ training samples, no adversarial perturbation with $\epsilon \leq 1.0$ could be found.

\begin{table}[htbp]
    \centering
    \small
    \setlength{\tabcolsep}{4pt} 
    \caption{
        Histogram of minimum $\epsilon$ required for PGD to generate adversarial examples on the training set.
    }
    \label{tab:epsilon-histogram}
    \begin{tabular}{cccc}
        \toprule
        \textbf{$\epsilon$} & \textbf{\#Samples} & \textbf{Correct} & \textbf{Flipped} \\
        \midrule
        $0.1$ & $63$  & $54$ & $5$ \\
        $0.2$ & $87$  & $81$ & $3$ \\
        $0.3$ & $92$  & $90$ & $0$ \\
        $0.4$ & $81$  & $77$ & $1$ \\
        $0.5$ & $87$  & $83$ & $0$ \\
        $0.6$ & $104$ & $103$ & $0$ \\
        $0.7$ & $88$  & $87$ & $0$ \\
        $0.8$ & $83$  & $79$ & $0$ \\
        $0.9$ & $71$  & $59$ & $0$ \\
        $1.0$ & $84$  & $78$ & $0$ \\
        \midrule
        \multicolumn{3}{r}{\textbf{No adversarial example $(\epsilon > 1.0)$:}} & \textbf{160} \\
        \bottomrule
    \end{tabular}
\end{table}

As shown in Table~\ref{tab:epsilon-histogram}, PGD produced successful
adversarial examples with \(\epsilon \leq 1.0\) for \(840\) out of \(1000\)
training samples. When evaluating the QNN on the resulting adversarially
perturbed inputs, the classification accuracy decreased from \(92.4\%\) on the
original inputs to \(1.07\%\) under adversarial perturbations.

\subsection{Stage 3: Adversarial Generation on Surrogates}\label{subsec:adversarial_examples_via_crosstalk}
In contrast to the first approach, we only consider those samples from the training data
that are correctly classified by the QNN when evaluated on the PennyLane statevector
simulator. Hence, we considered 885 out of 1000 samples for the surrogate-based
adversarial example generation. The following Table \ref{tab:epsilon-histogram_masked_samples} presents the distribution of the minimum $\epsilon$ values required to generate successful adversarial examples for the training set using masked PGD, as described in Section \ref{subsec:constraint_adversarial_example_generation}.

\begin{table}[htbp]
    \centering
        \caption{
        Histogram of minimum $\epsilon$ values needed for masked PGD to generate adversarial examples on the training set.
    }
    \label{tab:epsilon-histogram_masked_samples}
    \begin{tabular}{cccc}
        \toprule
        \textbf{Epsilon value} & \textbf{Number of samples} \\
        \midrule
        $0.1$ & $33$  \\
        $0.2$ & $28$  \\
        $0.3$ & $31$  \\
        $0.4$ & $50$  \\
        $0.5$ & $40$  \\
        $0.6$ & $28$  \\
        $0.7$ & $35$  \\
        $0.8$ & $28$   \\
        $0.9$ & $33$  \\
        $1.0$ & $32$ \\
        \midrule
        \multicolumn{3}{r}{\textbf{No adversarial example $(\epsilon > 1.0)$:}} & \textbf{547} \\
        \bottomrule
    \end{tabular}
\end{table}

As expected, it is quite hard to find adversarial examples given the relatively strong constraint that only inputs encoded in $R_Y$ gates can be manipulated. We evaluate three variants per selected training sample namely \textbf{clean} with the original input, \textbf{adversarial} with an adversarially perturbed input and \textbf{crosstalk} that approximates the adversarial sample through neighbor $R_Y$ rotations on adjacent qubits. Conceptually $x_{\mathrm{crosstalk}}$ is an approximation to $x_{\mathrm{adv}}$ with effective angles induced via rotations on the neighbors. From the $338$ training samples for which a successful adversarial example was
constructed on the PennyLane surrogate, we randomly select $50$ for execution on
the AQT Ibex hardware. The QNN qubits are placed at the given physical positions $[3,5,7,9]$ and neighbor $R_Y$ rotations are applied according to the left and right neighbors of each target qubit. All $150 = 3 * 50 $ circuits are transpiled with \verb|optimization_level = 0|. This layout implies that some disturbance qubits are shared between two QNN
qubits. For instance, a rotation applied to Q4 can induce crosstalk on both Q3
and Q5. This experiment is specific to the trapped-ion setting considered here. On AQT
Ibex, the qubits form a linear ion chain and single-qubit rotations are applied
by addressing individual ions, where residual illumination of neighbouring ions
can induce measurable crosstalk. Moreover, gates are executed sequentially on
this device, which makes it meaningful to place disturbance rotations at
specific points in the victim circuit.

We compute the $Z$ expectation values for the measured QNN qubits and assign the predicted class label as the argmax over the four expectation values. Accuracy is the fraction of correct predictions over the $50$ samples as summarized in Table \ref{tab:accuracy_crosstalk_masked_pgd_aqt}.

\begin{table}[htbp]
  \centering
    \caption{QNN accuracy on 50 randomly selected adversarially susceptible training samples.}
  \label{tab:accuracy_crosstalk_masked_pgd_aqt}
  \begin{tabular}{ l r }
    \toprule
    \textbf{Image} & \textbf{Accuracy} \\
    \midrule
    clean image        & 0.64 \\
    adversarial image  & 0.56 \\
    crosstalk image    & 0.68 \\
    \bottomrule
  \end{tabular}
\end{table}

These results are counterintuitive because the accuracy on the clean images is lower than expected and the overall differences, in terms of accuracy, between the three classes are smaller than anticipated. This can be explained by the fact that the selected samples were chosen based
on their behaviour under the PennyLane statevector simulator, whereas the
reported accuracies are obtained from AQT Ibex hardware. Consequently, samples that are correctly classified by
the simulator need not remain correctly classified under hardware and shot
noise. To quantify how well the AQT Ibex hardware results approximate the PennyLane statevector simulator we compare the four dimensional expectation value vectors for each sample. Let $e^{(A)}_i \in \mathbb{R}^4$ and $e^{(B)}_i \in \mathbb{R}^4$ denote the expectation values of two sources $A$ and $B$ for sample $i$, for example, $A$ refers to the PennyLane statevector simulator and $B$ to the AQT Ibex hardware.

The mean absolute error (MAE) for this sample $i$ is defined as
\begin{equation}\label{eq:mae}
  \Delta := \mathrm{MAE}_i(A,B)
  \;=\;
  \frac{1}{4} \sum_{k=1}^{4} \left| e^{(A)}_{i,k} - e^{(B)}_{i,k} \right|.
\end{equation}
Averaging over all $N = 50$ samples yields
\begin{equation}\label{eq:avg_mae}
  \overline{\mathrm{MAE}}(A,B)
  \;=\;
  \frac{1}{N} \sum_{i=1}^{N} \mathrm{MAE}_i(A,B).
\end{equation}

\begin{figure}[htbp]
  \centering
  \begin{minipage}[b]{0.48\textwidth}
    \centering
    \includegraphics[
        width=\linewidth,
        alt={Comparison of expectation values for unperturbed samples showing on the left two samples with the largest mean absolute error between AQT Ibex hardware results and the PennyLane statevector simulator, and on the right two samples with the smallest mean absolute error, illustrating strong input-dependent agreement or deviation between hardware and simulation}
    ]{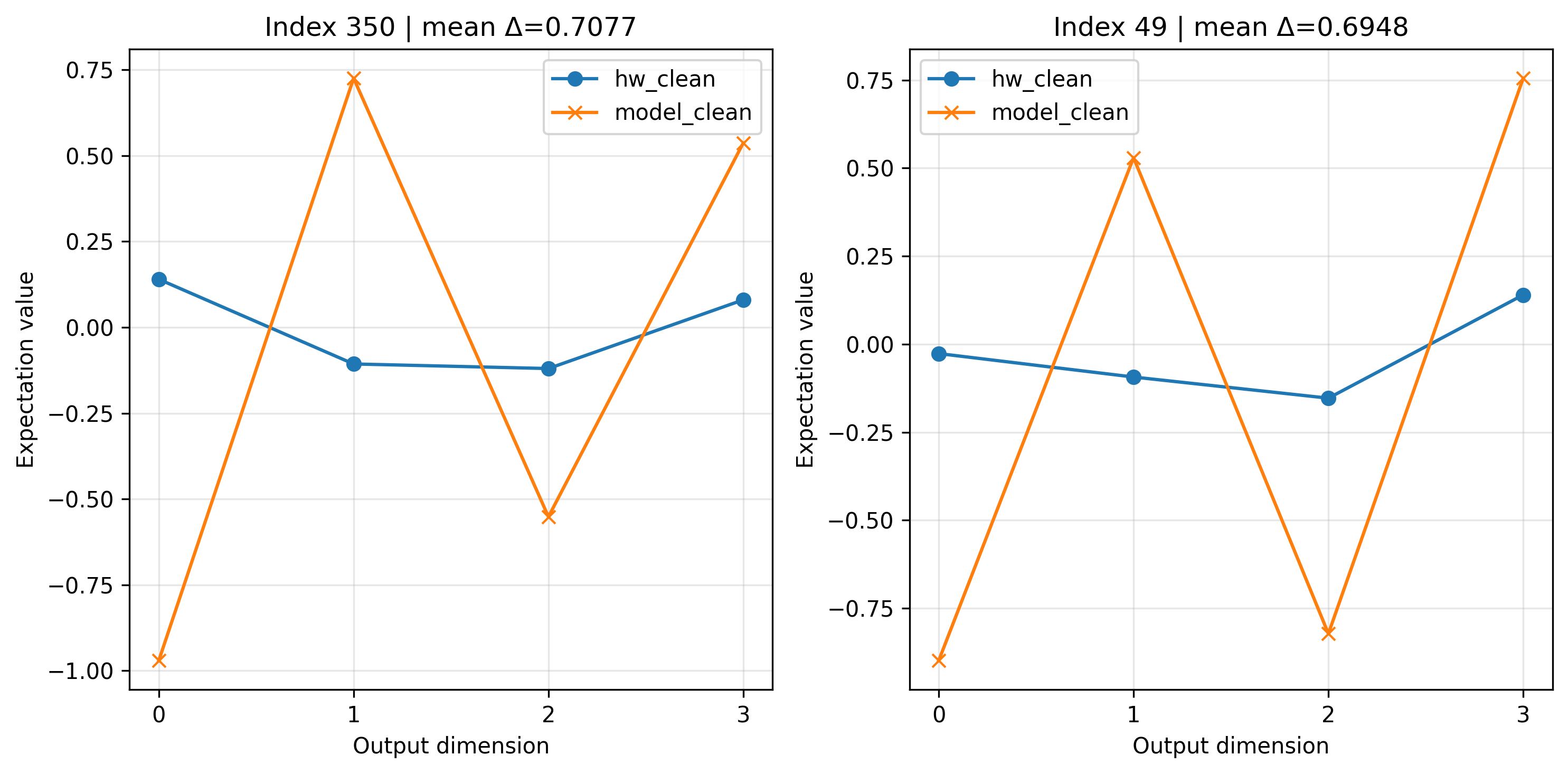}
  \end{minipage}
  \hfill
  \begin{minipage}[b]{0.48\textwidth}
    \centering
    \includegraphics[width=\linewidth]{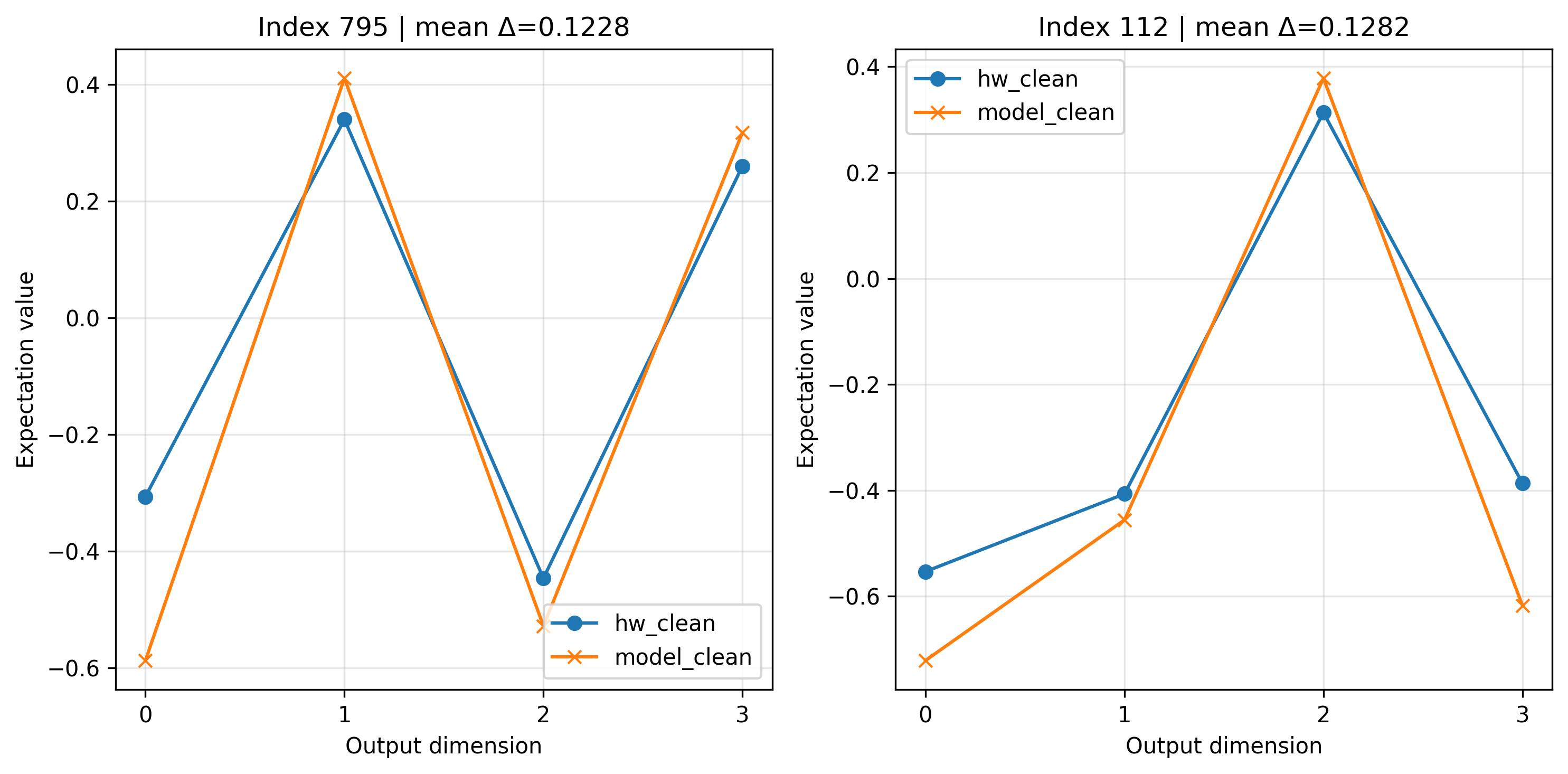}
  \end{minipage}
  \caption{Above Two samples with the largest mean absolute error. Below Two samples with the smallest mean absolute error.}
  \label{fig:mae_examples}
\end{figure}

To illustrate the spread of deviations across the inputs we inspect samples with particularly large and particularly small MAE between hardware and PennyLane statevector simulator expectation values. Figure~\ref{fig:mae_examples} shows the two unperturbed samples with the largest observed deviations and the two unperturbed samples with the smallest deviations. In each panel we plot the four expectation values from hardware and from the simulator.

We evaluate the mean absolute error between the full four-dimensional expectation
value vectors for all source pairs. Here, \texttt{hw\_clean} denotes the AQT Ibex
hardware expectation values for the original clean inputs, \texttt{hw\_adv} the
hardware expectation values for the adversarially perturbed inputs obtained by
masked PGD, and \texttt{hw\_crosstalk} the hardware expectation values for the
clean inputs executed with additional neighbour \(R_Y\) rotations designed to
induce an adversarial crosstalk pattern. Analogously, \texttt{model\_clean} and
\texttt{model\_adv} denote the PennyLane statevector simulator expectation values
for the clean and adversarial inputs, respectively.

Using the definition of the average mean absolute error in \cref{eq:avg_mae},
Table~\ref{tab:mae_pairs} reports \(\overline{\mathrm{MAE}}(A,B)\) for all pairs
\((A,B)\), averaged over all fifty samples. In the extended
report~\cite{bsi_qml_esa_2026}, we also consider confidence-based sample
selection, focusing on inputs for which the victim QNN exhibits either the
highest or the lowest predictive certainty when evaluated using the PennyLane
statevector simulator.

\begin{table}[htbp]
    \centering
        \caption{Mean absolute error between expectation value vectors for all pairs of hardware and simulator sources, averaged over fifty samples.}
    \label{tab:mae_pairs}
    \small
    \begin{tabular}{@{}lc@{}}
        \toprule
        \textbf{Pair} & \textbf{$\overline{\mathrm{MAE}}$} \\
        \midrule
        hw\_adv       vs hw\_crosstalk & 0.1236 \\
        hw\_clean     vs hw\_crosstalk & 0.1270 \\
        hw\_clean     vs hw\_adv       & 0.1452 \\
        model\_clean  vs model\_adv    & 0.1682 \\
        hw\_adv       vs model\_adv    & 0.2532 \\
        hw\_clean     vs model\_adv    & 0.2745 \\
        hw\_crosstalk vs model\_adv    & 0.2831 \\
        hw\_clean     vs model\_clean  & 0.3086 \\
        hw\_crosstalk vs model\_clean  & 0.3294 \\
        hw\_adv       vs model\_clean  & 0.3534 \\
        \bottomrule
    \end{tabular}
\end{table}

It is encouraging that the smallest mean absolute error is observed for the pair
\texttt{hw\_adv} versus \texttt{hw\_crosstalk}. 

\newpage This indicates that the crosstalk
construction indeed moves the hardware evaluation of the clean inputs closer to
the adversarial hardware behavior, which is precisely the intended purpose of
the \texttt{hw\_crosstalk} configuration.

\section{Discussion}\label{sec:discussion}
\subsection{Key Findings}
In this paper, we executed a multi-stage attack on AQT trapped-ion hardware.
During the reconnaissance stage, the power-trace attack enabled the most direct
and reliable reconstruction of the victim QNN architecture. Without any
circuit-level optimization, the comparison with benchmark circuits yielded an
unambiguous result, allowing us to infer the complete QNN architecture. A broader
comparison with crosstalk-based and timing-based reconnaissance attacks is
provided in the QML-ESA report~\cite{bsi_qml_esa_2026}, where power traces emerge
as the strongest among the evaluated approaches. The report further demonstrates
that the weights and bias parameters encoded in the $R_Y$ gates can be reverse
engineered. This, however, is generally not possible for parameters encoded in $R_Z$ gates, as these gates are implemented virtually. Furthermore, the adversarial-example results in Section~\ref{subsec:adversarial_examples}
show that the trained QNN is highly vulnerable to input-space perturbations:
PGD succeeds on a large majority of the analyzed training samples and causes a
near-complete collapse in classification accuracy under adversarial perturbation.

In addition, in the report~\cite{bsi_qml_esa_2026}, we analyze transfer attacks and show that adversarial examples constructed under statevector simulation partially retain their effectiveness when executed on real superconducting and trapped-ion hardware. While backend noise reduces the overall misclassification rates, the decision-boundary shifts induced by the adversarial perturbations persist across different hardware platforms.

The approximation of Adversarial examples via Crosstalk proved successful in the sense that the $\overline{\mathrm{MAE}}$ is minimized between the adversarial examples and the original samples perturbed through crosstalk, as outlined in Table \ref{tab:mae_pairs}. A potential extension of the crosstalk-based adversarial example construction concerns the choice of qubit layout. In the current experiments, the victim QNN qubits reside on the physical qubits $[3,5,7,9]$, which implies that inducing a disturbance on one target qubit necessarily requires applying rotation gates to its physical neighbours. For example, attempting to perturb qubit Q3 via its neighbours Q2 and Q4 inevitably affects qubit Q5 as well, since the applied $R_Y$ rotations on Q4 induce crosstalk on both adjacent qubits. It would therefore be interesting to explore whether alternative, intentionally non-standard layouts, such as placing the QNN qubits on positions $[2,5,8,11]$, could strengthen the attack. In such a configuration, the disturbance qubits $[1,4,7,10]$ would interface with the QNN without unintentionally influencing additional neighbours. However, it must be emphasised that such layouts are not typical and should be regarded primarily as an academic construct aimed at probing the limits of crosstalk-induced adversarial manipulation.

\subsection{Defensive Strategies}
\subsubsection{Power Traces}
To mitigate the risk of power side-channel attacks  several defensive strategies have been proposed in the literature. According to \cite{erata_quantum_2024}, the following approaches represent the most immediate and effective lines of defense:

\begin{itemize}
    \item \textbf{Decoy Pulses:} Executing calculation-independent gates that do not contribute to the original circuit but interfere with the power trace profile, which makes circuit reconstruction computationally more expensive (increasing number of variables) and noisier (more influence of parallel gates)
    
    \item \textbf{Power Randomization:} Introducing random variations in power consumption, such as randomized pulse timing, or stochastic load activation

    \item \textbf{Constant-Power Operation:} Designing the arbitrary waveform generators (AWGs) or FPGA-based controllers to maintain a constant power profile regardless of whether quantum pulses are being generated or idle

\end{itemize}

These countermeasures, inspired by classical side-channel defenses, require adaptation to the control architectures of quantum computers \cite{erata_quantum_2024}.

\subsubsection{Adversarial Examples}

A fundamental prerequisite for robustness in machine learning systems is a thorough understanding of the model's behavior under adversarial conditions and the integration of corresponding defense mechanisms during training. Quantum neural networks, similar to their classical counterparts, can benefit from two key strategies: adversarial training and regularization. Adversarial training involves deliberately exposing the model to perturbed inputs during training, thereby enabling it to generalize better under adversarial stress. This method is conceptually simple and has proven effective in increasing robustness in both quantum and classical domains. More rigorous claims of robustness can be made when restricting the Lipschitz bound of the model’s encoding Hamiltonians, via Lipschitz regularization of the quantum model. Lipschitz bounds quantify how much a model’s output can change concerning small changes in its input. Lower Lipschitz bounds signify that a model is less sensitive to such changes, making it inherently more robust against adversarial attacks \cite{berberich_training_2023,wendlinger_comparative_2024}.

\subsubsection{Adversarial Examples via Crosstalk}
The adversarial examples discussed in Section~\ref{subsec:adversarial_examples_via_crosstalk}, although restricted to perturbations on input components encoded in $R_Y$ rotations, constitute genuine adversarial manipulations of the QNN. For this reason, the defensive mechanisms discussed for adversarial examples, apply equally in the present context. Both approaches reduce the model’s sensitivity to local input variations and thus mitigate the effectiveness of adversarial examples. In addition to model-level defenses, effective protection against crosstalk-based adversarial behavior also requires system-level measures. The most direct and robust defence is to avoid situations in which multiple users share the same quantum processor at the same time.

\newpage If no shared-execution scenario is permitted, an adversary cannot place rotation gates on neighboring qubits and therefore cannot induce effective angles on the victim circuit. Preventing shared execution removes this attack vector entirely. A further defensive consideration arises from the fact that an attacker must possess sufficiently accurate knowledge of the execution time of the victim circuit in order to place disturbance gates at the correct positions in time. For this reason, defensive mechanisms against model-stealing attacks are relevant.

\subsection{Ethics}
All experiments presented in this paper were conducted in controlled research settings and exclusively on quantum hardware and simulators to which the authors had legitimate access. No production systems, customer workloads, or confidential user data were targeted or affected. The primary motivation of this work is to contribute to the secure design and deployment of quantum computing and QML systems. By systematically exposing attack vectors and their interdependencies, the results aim to inform hardware providers, cloud operators, and QML practitioners about potential risks that may not be apparent from isolated threat analyses.

\section{Conclusion}\label{sec:conclusion}
In this work, we presented an end-to-end, multi-stage kill-chain attack against a QNN executed on real trapped-ion quantum hardware. By chaining side-channel reconnaissance, adversarial example generation, and hardware-induced crosstalk manipulation, we demonstrated that vulnerabilities across different abstraction layers of quantum machine learning systems can be combined into a coherent attack pipeline. Our results show that information leaked through side channels, in particular power traces, enables the inference of the victim QNN’s architectural structure, which in turn allows the attacker to construct targeted noise-injection attacks. Overall, these findings highlight the necessity of analyzing QML security from a system-level perspective, rather than treating noise, crosstalk, and adversarial robustness as independent phenomena.
\section*{Acknowledgment}
This research is part of the \emph{QML-ESA} (Extended Security Analysis of Quantum Machine Learning) project, which is funded by the German Federal Office for Information Security (BSI)~\cite{bsi_qml_esa_2026}.
\newpage 
\appendices
\section{Superconducting Experiments}

\subsection{Stage-by-Stage Outcomes}
As in the trapped-ion case, we explored side-channel attacks based on crosstalk, timing, and power traces. Since the power-trace attack is the most promising among these, it is discussed in the appendix. With respect to noise injection, we briefly address the Active SWAP attack in the appendix, while a substantially more detailed treatment of this attack, alongside other attack vectors, is provided in the extended report~\cite{bsi_qml_esa_2026}.
\subsubsection{Power Trace Attack}
\begin{figure}[htbp]
  \centering
  \begin{minipage}{0.2\textwidth}
    \centering
    \includegraphics[
      width=0.95\linewidth,
      alt={On the left: Toy circuit consisting of $\sqrt{X}$ and X gate on the first qubit, X and $\sqrt{X}$ on the second qubit, followed by a CNOT gate.}
    ]{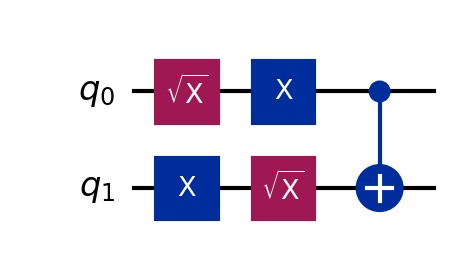}
  \end{minipage}\hfill
  \begin{minipage}{0.3\textwidth}
    \centering
      \caption{Toy-CX test circuit: original (top) and after transpilation (bottom).}
  \label{fig:total_power_attack_toy_cx_circuits}
    \includegraphics[
      width=0.95\linewidth,
      alt={On the right: Same circuit after transpilation, where single-qubit gates are decomposed into RZ and $\sqrt{X}$ gates.}
    ]{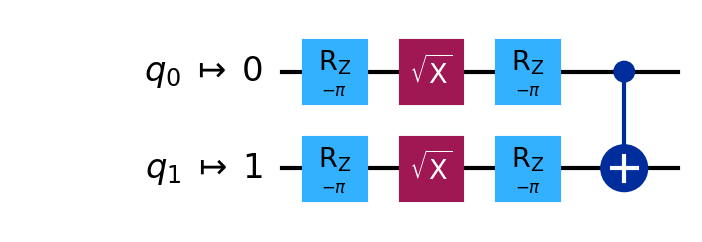}
  \end{minipage}
\end{figure}

As in the trapped-ion case, we do not have access to actual power traces, we base all experiments on power traces computed from pulse schedules. More specifically, we generate pulse schedules and power traces from IBM's simulated pulse backend \emph{Fake7QPulseV1}\footnote{\url{https://docs.quantum.ibm.com/api/qiskit/1.0/qiskit.providers.fake_provider.Fake7QPulseV1}}, which provides an interface to the pulse-level information for each qubit in the circuit. The target circuit and its transpiled version is shown in \Cref{fig:total_power_attack_toy_cx_circuits}. This toy circuit consists of two qubits and five gates: Two single-qubit gates on each qubit, namely a $\sqrt{X}$ gate followed by an $X$ gate and their permuted version, and one CNOT between the first and the second qubit. During transpilation, this circuit is decomposed into an $R_Z$ - $\sqrt{X}$ - $R_Z$ gate combination on both the first and the second qubit, followed by the CNOT gate. Notably, $R_Z$ gates are implemented as \emph{virtual gates} on the IBM superconducting hardware, which leaves no measurable effect on the power trace.

\begin{figure}[htbp]
    \centering
        \caption{Pulse Schedule for Toy-CX Test Circuit}
    \label{fig:total_power_attack_toy_cx_schedule}
\includegraphics[width=0.80\linewidth,
    alt={per-channel pulse schedule for the toy-CX circuit. The pulses show single-qubit and two-qubit gate envelopes. On this per-channel basis, one can easily classify the original gates from pulse shapes.}]{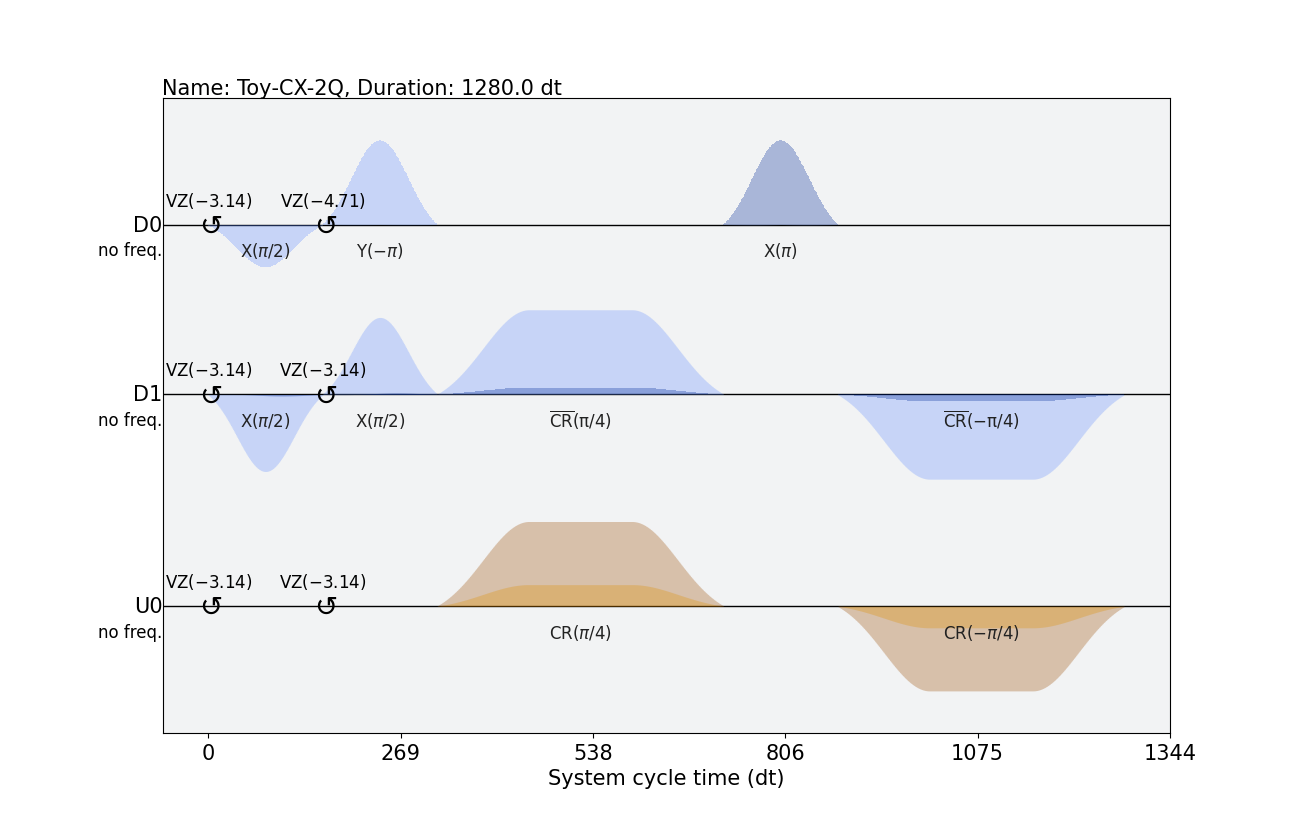}
\end{figure}

For the first experiment, we use the pulse simulator to generate noise-free per-channel power traces. Given the toy circuit as described, the resulting power trace schedule is depicted in \Cref{fig:total_power_attack_toy_cx_schedule}. 

As described in the report~\cite{bsi_qml_esa_2026}, the objective of an attacker can be stated as 
\begin{align*}
    A_{PL} &= \arg \min_{A'_{PL}} \sum_{c \in C} d\Bigl\{ \text{Power}_c[A'_{PL}](x),\, v_c(x) \Bigr\},
\end{align*}
where a simple per-channel analysis of the power trace signatures of the target and benchmark circuits is enough to determine the most likely candidate.  Some details of the template scoring are shown in Table~\ref{tab:per_channel_reconstruction_jsd}. For each relevant time step the lowest score is chosen and while  the resulting channel occupation constraints are tracked. This results in some per-time-step ranking lists being shorter than others as in the case of $t=0$ for the \emph{second} qubit.

\begin{table}[h!]
    \centering
        \caption{Jensen–Shannon distance scoring and ranking of gate templates at key times steps (0, 160)}
    \label{tab:per_channel_reconstruction_jsd}
    \begin{tabular}{lllll}
        \toprule
        Rank & $t$ & Gate & Qubit & $\sqrt{\text{JSD}}$  \\
        \midrule
		\textbf{1} & 0	 & SX & 0	 	& \textbf{0.00000000}	 \\
		2 & 0	 & SX & 1	 	& 0.00000000	 \\
		3 & 0	 & X & 0	 	& 0.00000582	 \\
		4 & 0	 & X & 1	 	& 0.00027877	 \\
		5 & 0	 & CX & 0,1	 	& 1.84281222	 \\
		6 & 0	 & CX & 1,0	 	& 1.89385275	 \\
          \midrule
		\textbf{1} & 0	 & SX & 1	 	& \textbf{0.00000000}	 \\
		2 & 0	 & X & 1	 	& 0.00027877	 \\
          \midrule
		\textbf{1} & 160	 & CX & 0,1	 	& \textbf{0.00000000}	 \\
		2 & 160	 & SX & 1	 	& 0.00000000	 \\
		3 & 160	 & X & 0	 	& 0.00000000	 \\
		4 & 160	 & SX & 0	 	& 0.00000582	 \\
		5 & 160	 & X & 1	 	& 0.00027877	 \\
        \bottomrule
    \end{tabular}
\end{table}

From this table the schedule can be easily reconstructed as shown in Table~\ref{tab:perchannel_power_attack_reconstructed_schedule}. As noted, the reconstruction results only include SX gates since we cannot measure the effect of virtual $R_Z$ gates in the power trace.

\begin{table}[h!]
    \centering
        \caption{Reconstructed Gate Schedule for Toy-CX Circuit using a per-channel power traces.}
    \label{tab:perchannel_power_attack_reconstructed_schedule}
    \begin{tabular}{ccc}
    \toprule
        $t_s$   & gate  &   qubit\\
    \midrule
        0       & SX    &   0\\
        0       & SX    &   1\\
        160     & CX    &   0,1\\
    \bottomrule
    \end{tabular}
\end{table}

\begin{figure}[htbp]
    \centering
\includegraphics[width=0.90\linewidth,
alt={Per-channel and total power trace pulse lines overlayed. From the total power trace, it is harder to find out which gates were executed due to interferences in the traces.}]{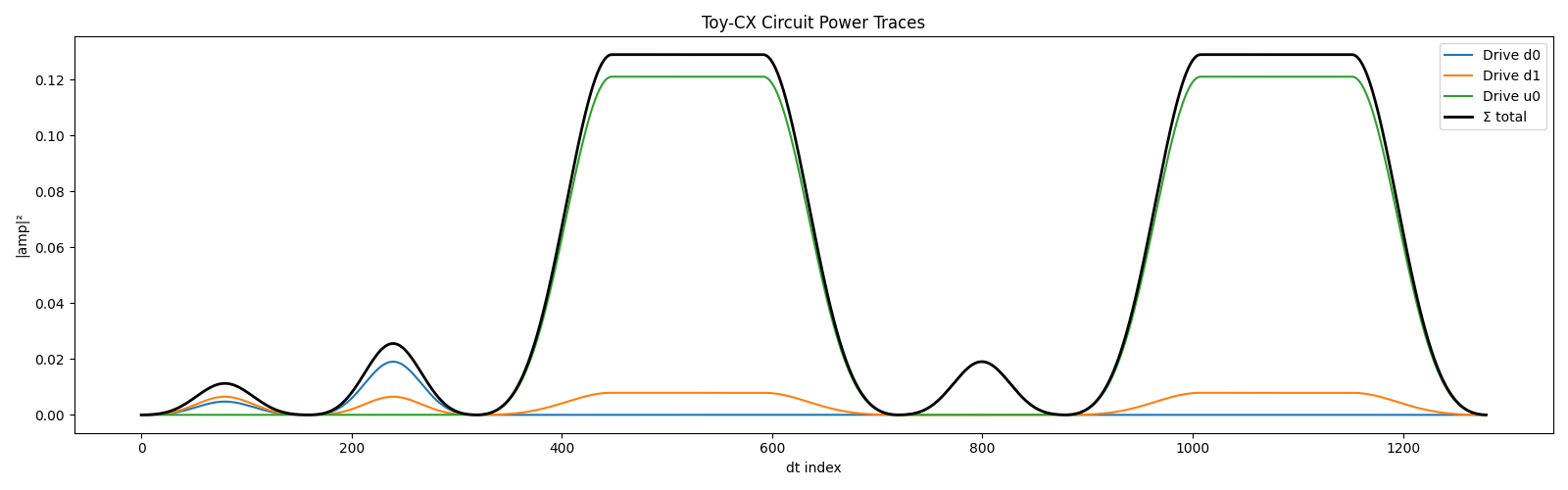}
    \caption{Per-Channel and Total Power Trace for Toy-CX Test Circuit}
    \label{fig:total_power_attack_toy_cx_trace}
\end{figure}

The situation changes when the information leakage is restricted to the combined power trace, where the sum of individual gates makes it hard for an attacker to infer which gates were executed at which timestep on which qubit. This issue becomes clear when looking at \Cref{fig:total_power_attack_toy_cx_schedule} and \Cref{fig:total_power_attack_toy_cx_trace} as we can directly infer which gate is executed on which qubit in the per-channel power trace schedule; however, for the summed total power trace, this is not trivially possible as permuted gate-to-qubit assignments may lead to similar total power traces.

In the QML-ESA report~\cite{bsi_qml_esa_2026}, we discuss how to tackle this challenge.

\subsubsection{Active SWAP Attack}
In the Active SWAP attack, to our knowledge first introduced by \cite{lee_swap_2025}, the attacker executes a single CNOT gate concurrently with the victim’s circuit on a cloud-based, multi-tenant quantum system in order to induce errors in the victim’s output. The purpose of this attack is to disturb the computation of the victim QNN and thereby reduce the accuracy of the model. Based on the QNN layout identified by the attacker during the reconnaissance phase, a benchmark circuit is constructed by applying a Hadamard gate to each QNN qubit. For each candidate pair of qubits, the attacker executes a fixed number of shots while concurrently applying a single CNOT gate together with this benchmark circuit. The resulting bitstring distribution is then compared to the uniform distribution, which is the expected outcome of the benchmark circuit in the absence of the disturbing CNOT gate, using the trace distance between the two probability distributions. The trace distance is defined as
    \begin{equation}\label{eq:trace_distance}
    D(P, Q) \equiv D(p_x, q_x) = \frac{1}{2} \sum_x |p_x - q_x|,
    \end{equation}
    where \( P = \{ p_x \}_x \) and \( Q = \{ q_x \}_x \) are two probability distributions over the same index set. Qubit pairs that maximize this trace distance are selected, as they are expected to induce the strongest crosstalk effects and are therefore used to interfere with the execution of the victim QNN. As discussed in depth in the report~\cite{bsi_qml_esa_2026}, it is appropriate to consider a benchmark circuit that does not require any SWAP operations in order to be executed on the device.

\begin{figure}[htbp]
    \centering
    \includegraphics[width=0.72\linewidth]{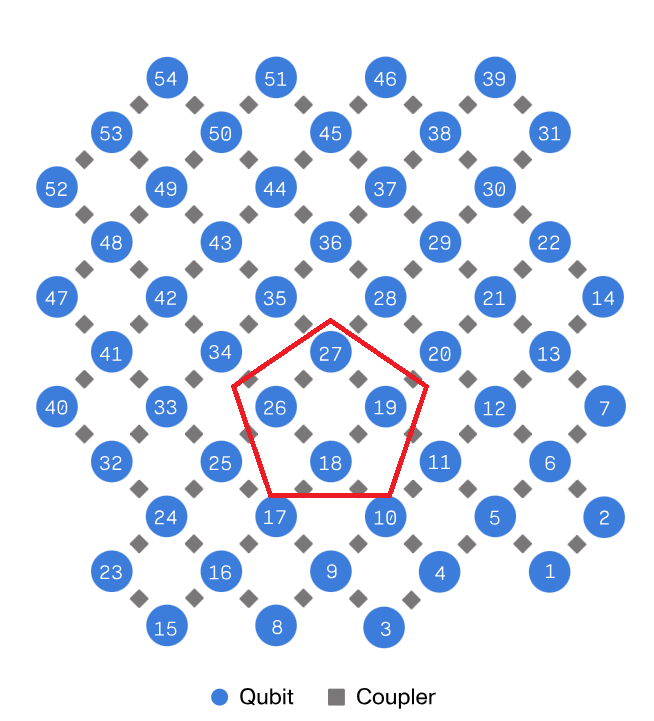}
    \caption{Physical qubit layout of the 54-qubit IQM Emerald device. The red outline indicates the qubits used for the victim QNN in the Active SWAP experiment.}
    \label{fig:iqm_emerald_qubit_placement}
\end{figure}

The experiment is performed on the 54-qubit IQM Emerald superconducting device.
The physical qubit layout used for this experiment is shown in
Figure~\ref{fig:iqm_emerald_qubit_placement}. The victim QNN is placed on the
physical qubits \([18,19,26,27]\), while candidate disturbance CNOTs are swept
across the remaining device.

We fix qubit 1 and sweep the second qubit \(j\) across the device. For each candidate index \(j\), we construct a quantum circuit by applying a CNOT from qubit 1 to qubit \(j\). Now we apply a Hadamard gate on each of the qubits \([18,19,26,27]\) and  measure those four target qubits. We record the resulting bitstring distribution and compute its trace distance to the uniform distribution. We rank the CNOT pairs by their trace distance to the uniform distribution and summarize the top five in Table \ref{tab:top5_cnot_trace_distance_uniform_distribution_control_qubit_one}. 
\newpage 
\begin{table}[htpb]
  \centering
    \caption{Top five CNOT pairs by trace distance (control qubit 1).}
\label{tab:top5_cnot_trace_distance_uniform_distribution_control_qubit_one}
  \begin{tabular}{l r}
    \toprule
    \textbf{CNOT Pair} & \textbf{Trace Distance} \\
    \midrule
    (1, 42) & 0.1018 \\
    (1, 43) & 0.0859 \\
    (1, 34) & 0.0857 \\
    (1, 38) & 0.0825 \\
    (1, 36) & 0.0806 \\
    \bottomrule
  \end{tabular}
\end{table}

In a second run, we fix qubit 54 and sweep the second qubit \(j\) across the device. For each candidate index \(j\), we construct a quantum circuit by applying a CNOT from qubit 54 to qubit \(j\), apply the Hadamard gates on physical qubits \([18,19,26,27]\) and then measure those four target qubits. We rank the CNOT pairs by their trace distance to the uniform distribution and summarize the top five in Table \ref{tab:top5_cnot_trace_distance_uniform_distribution_control_qubit_fiftyfour}.
\begin{table}[htpb]
  \centering
    \caption{Top five CNOT pairs by trace distance (control qubit 54).}
  \label{tab:top5_cnot_trace_distance_uniform_distribution_control_qubit_fiftyfour}
  \begin{tabular}{l r}
    \toprule
     \textbf{CNOT Pair} & \textbf{Trace Distance} \\
    \midrule
    (54, 3)  & 0.1472 \\
    (54, 4)  & 0.1084 \\
    (54, 9)  & 0.0918 \\
    (54, 15) & 0.0708 \\
    (54, 10)  & 0.0706 \\
    \bottomrule
  \end{tabular}
\end{table}

\begin{figure}[htbp]
  \centering
  \begin{minipage}{0.35\textwidth}
    \centering
    \includegraphics[width=\textwidth]{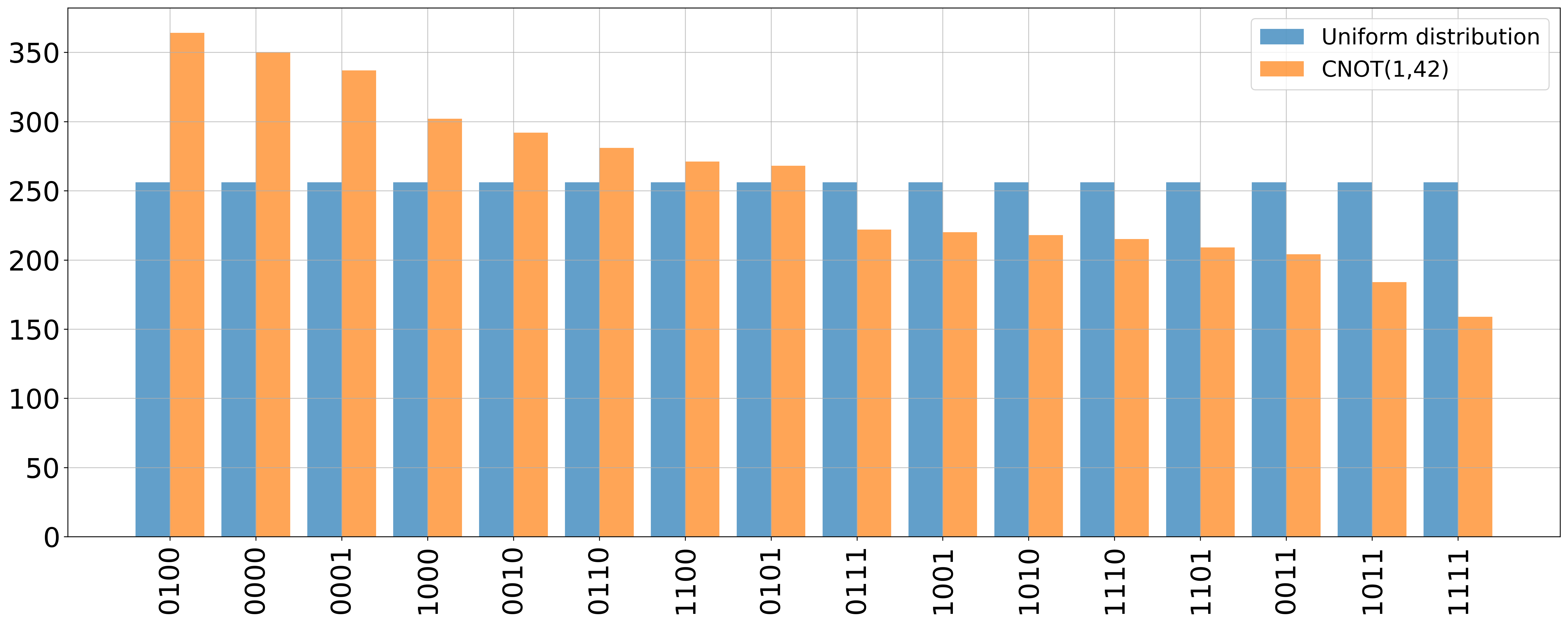}
  \end{minipage}\hfill
  \begin{minipage}{0.35\textwidth}
    \centering
    \includegraphics[width=\textwidth]{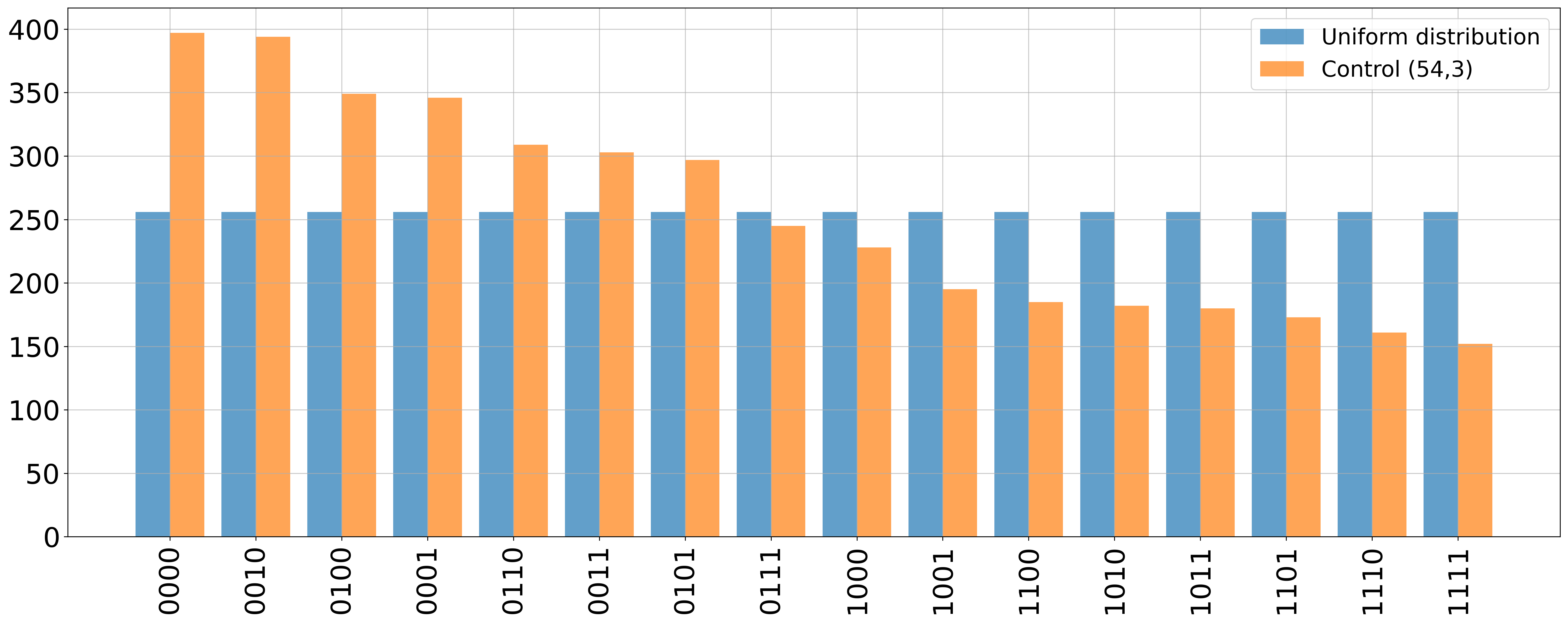}
  \end{minipage}
  \caption{Bitstring distribution comparison between the uniform distribution and disturbance. \emph{left} disturbance on $(1,42)$ \emph{right} disturbance on $(54,3)$ }
  \label{fig:trace_distances_to_uniform_distribution}
\end{figure}

In the following plot \ref{fig:trace_distances_to_uniform_distribution} we compare the bitstring distributions of the candidate CNOT pairs $[(1,42), (54,3)]$ with the uniform distribution.

Those plots suggest that applying a CNOT on one of those pairs induces a significant amount of noise in the victim QNN, since they induce a significantly different bitstring distribution compared to the expected uniform distribution

When we calculate the accuracy of the QNN on 100 test images with 25 images per class, we get the following \ref{tab:accuracy_drop__multiple_CNOTs_iqm_emerald_new_approach}:  

\begin{table}[htbp]
  \centering
    \caption{QNN accuracy on 100 test images for each disturbance.}
  \label{tab:accuracy_drop__multiple_CNOTs_iqm_emerald_new_approach}
  \begin{tabular}{ l r }
    \toprule
    \textbf{CNOT Pairs} & \textbf{Accuracy} \\
    \midrule
    no disturbance        & 0.8 \\
    (1, 42)                & 0.87 \\
    (54, 3)              & 0.88 \\
    (1, 42), (54, 3)      & 0.75 \\
    \bottomrule
  \end{tabular}
\end{table}

It is interesting to note that applying either CNOT(1, 42) or CNOT(54, 3) individually leads to an increase in the model's accuracy compared to the undisturbed case. However, when both CNOTs are applied simultaneously, the accuracy of the model actually decreases. This should not be interpreted as a systematic accuracy improvement caused by
the disturbance CNOTs. Rather, the disturbances can have a counterintuitive
stabilising effect on this particular set of test images by reducing specific
misclassification patterns. At the same time, the combined disturbance changes
the misclassification behaviour more strongly and shifts errors from one class
to another, which is consistent with the observed decrease in overall accuracy.

\begin{figure}[htbp]
  \centering
  \setlength{\tabcolsep}{2pt}

  \begin{tabular}{@{}cc@{}}
    \includegraphics[width=0.49\linewidth]{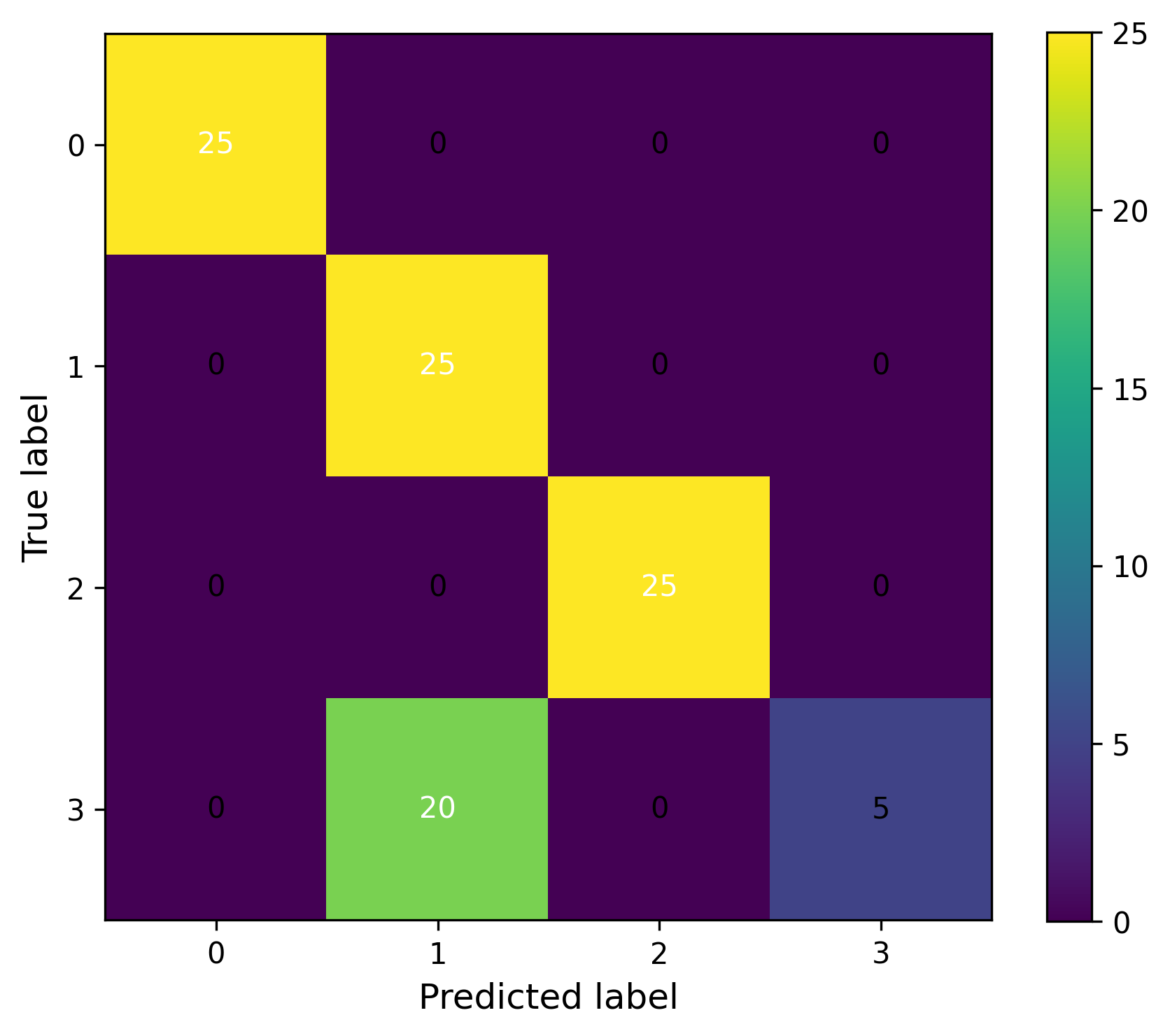} &
    \includegraphics[width=0.49\linewidth]{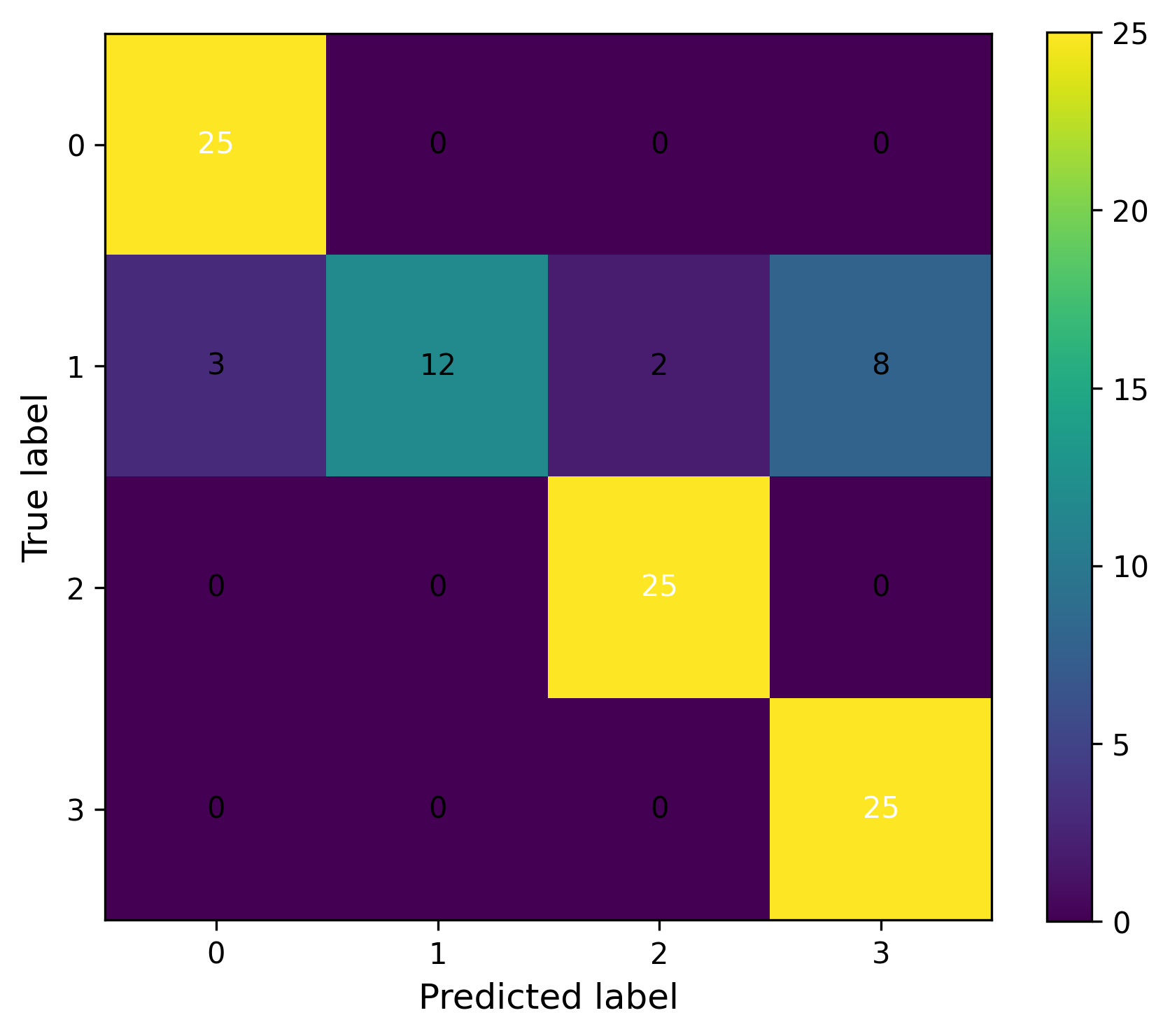} \\
    \includegraphics[width=0.49\linewidth]{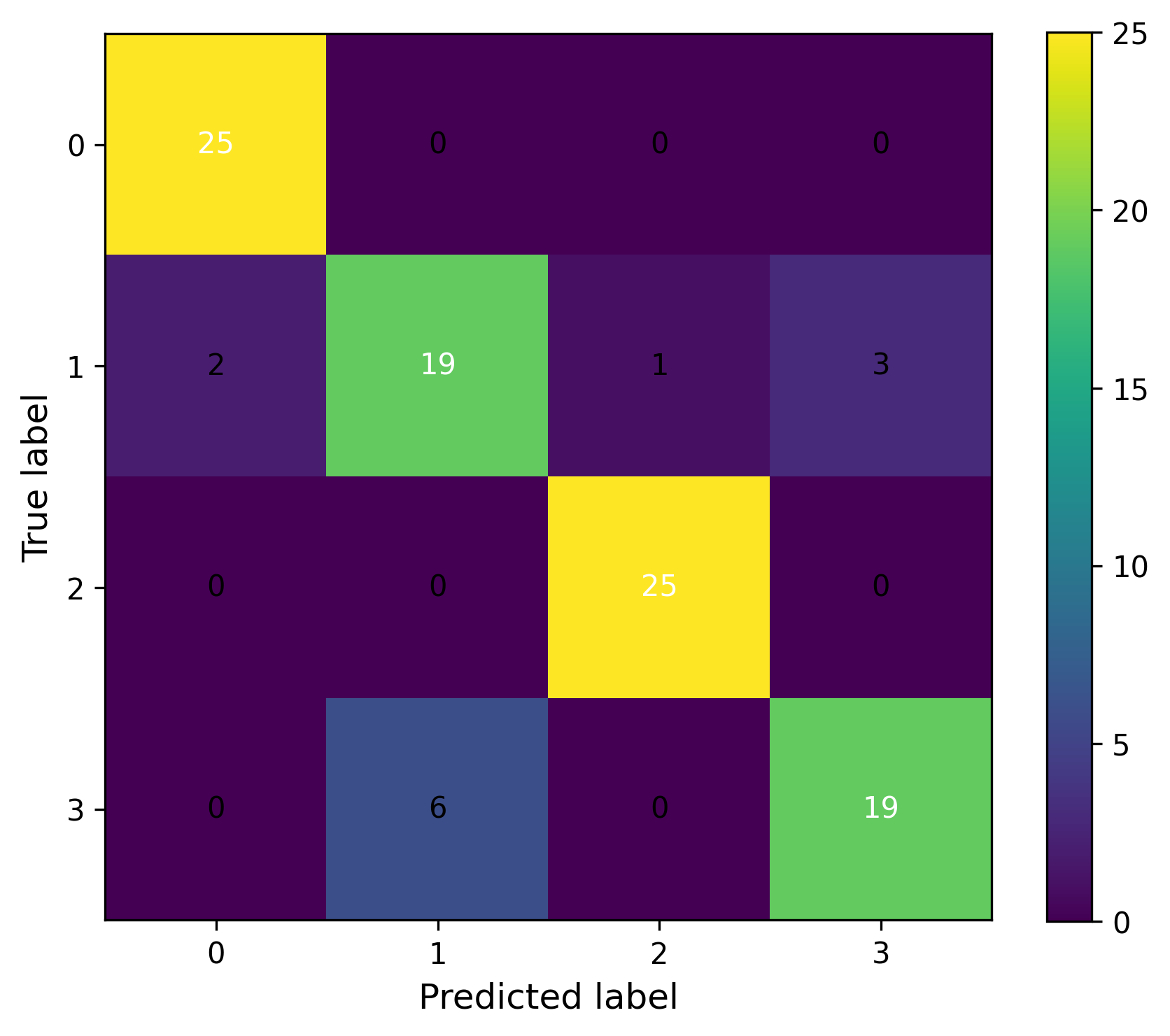} &
    \includegraphics[width=0.49\linewidth]{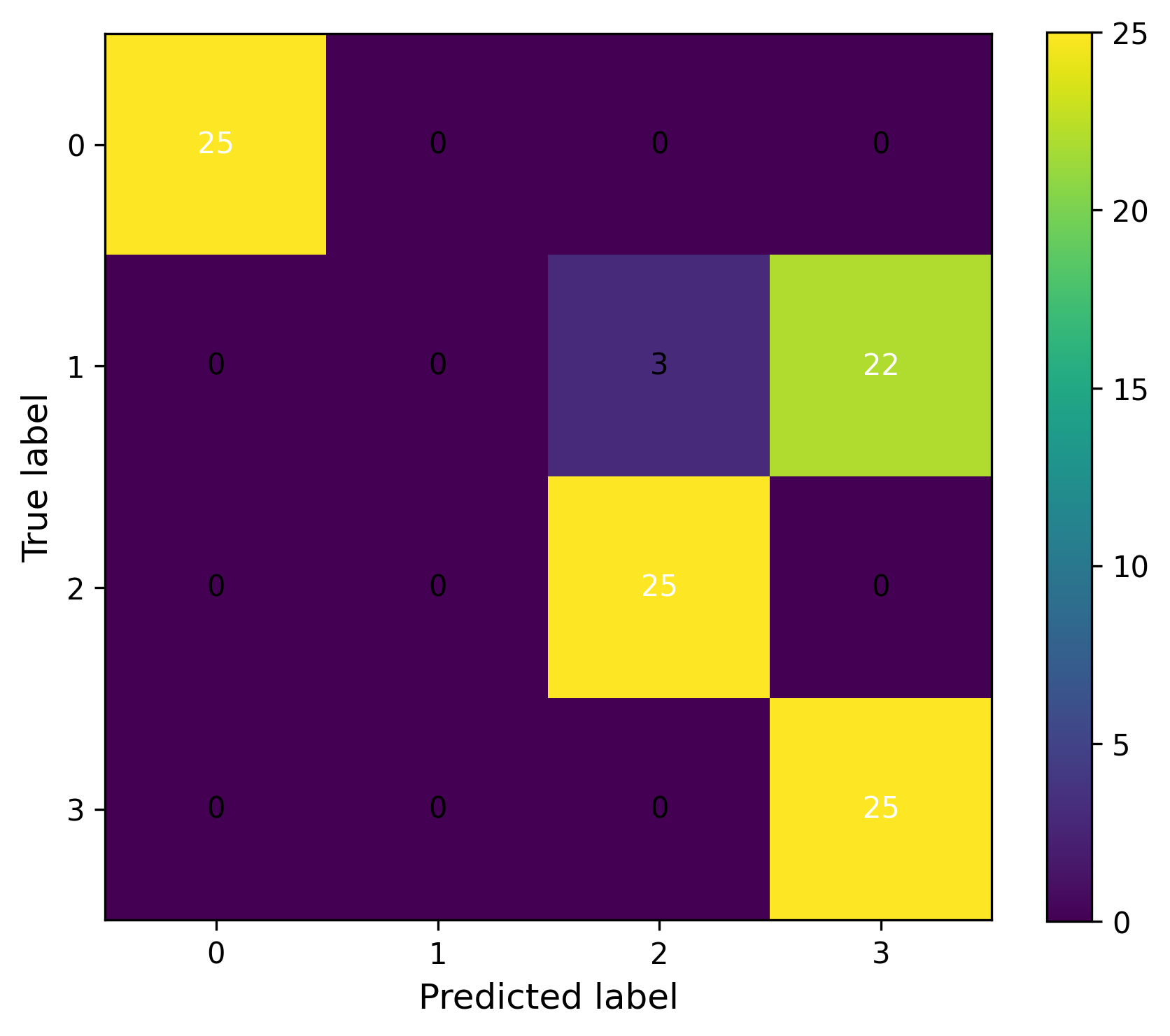}
  \end{tabular}

  \caption{Confusion matrices for the baseline execution (top left), disturbances on qubit pairs $(1,42)$ (top right) and $(54,3)$ (bottom left), and the combined disturbance on $(1,42)$ and $(54,3)$ (bottom right).}
  \label{fig:confusion_matrices_general_fault_injection}
\end{figure}

A clear view of the classification behaviour under each disturbance scenario is obtained by examining the corresponding confusion matrices. Figure~\ref{fig:confusion_matrices_general_fault_injection} compares the confusion matrices for the baseline execution, the two individual disturbances \((1,42)\) and \((54,3)\), and the combined disturbance. 

In the report~\cite{bsi_qml_esa_2026}, we further discuss a topology-aware Active SWAP attack, where the idea is to take advantage of the topology of the quantum hardware to find more suitable CNOT gate candidates for influencing the victim circuit.

\section{AQT Crosstalk Experiment}\label{sec:aqt_crosstalk_experiment}
We aim to explore the concept of crosstalk signature analysis on the AQT Ibex 12-qubit device. As discussed in Section \ref{subsec:crosstalk_mechanisms}, on trapped-ion devices, crosstalk is primarily induced by single-qubit rotations. We began with a proof-of-concept attack, in which the victim circuit consists of four qubits located on ions 3, 5, 7, and 9. On each qubit, we apply ten $R_Y$ gates with an angle of $\pi/2$, inserting a barrier after each gate to prevent the compiler from combining them into a single $R_Y$ rotation. We want to collect a crosstalk signature by measuring the neighboring ions $[2,4,6,8,10]$. 

In the following Figure \ref{fig:crosstalk_signature_analysis_aqt_ibex_victim_circuit}, the victim circuit is illustrated:

\begin{figure}[htbp]
    \centering
    \includegraphics[
        width=1\linewidth,
        alt={Conceptual circuit diagram illustrating the victim setup for the crosstalk signature attack on the AQT Ibex device, showing four victim qubits on ions 3, 5, 7, and 9 with repeated single-qubit $R_Y$ rotations separated by barriers, and neighboring ions used as measurement qubits to capture induced crosstalk}
    ]{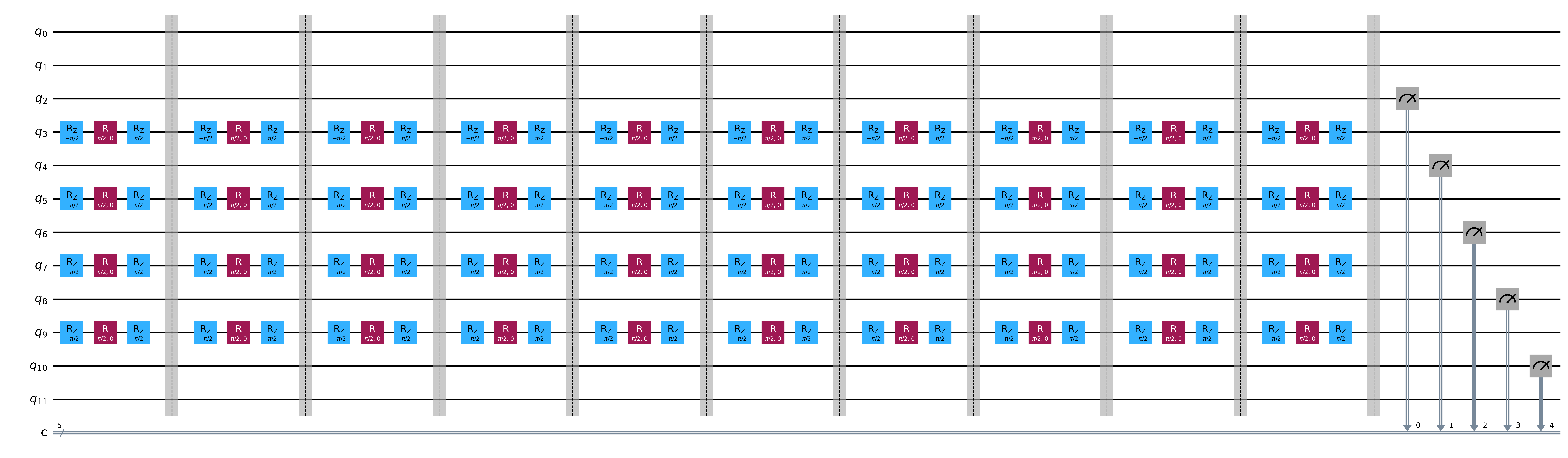}
    \caption{A conceptual overview of the victim circuit.}
    \label{fig:crosstalk_signature_analysis_aqt_ibex_victim_circuit}
\end{figure}

We performed the maximum of 2000 shots per job and obtained the bitstring distribution shown in Table~\ref{tab:aqt_ibex_victim_circuit_bitstrings}.

\begin{table}[h!]
    \centering
        \caption{Measured bitstring distribution over 2000 shots on the AQT Ibex device.}
    \label{tab:aqt_ibex_victim_circuit_bitstrings}
    \begin{tabular}{cc}
        \toprule
        \textbf{Bitstring} & \textbf{Counts} \\
        \midrule
        00000 & 1849 \\
        00010 & 44 \\
        00100 & 51 \\
        01000 & 42 \\
        00001 & 6 \\
        10000 & 6 \\
        01100 & 1 \\
        00110 & 1 \\
        \bottomrule
    \end{tabular}
\end{table}

The table clearly shows that crosstalk is induced on the neighboring ions of the AQT Ibex device. In a noise-free scenario, we would expect to observe the bitstring 00000 in all 2000 shots. However, it is notable that qubits 4, 6, and 8 are measured in the 1 state significantly more often than the edge qubits 2 and 10. This observation can be explained by the fact that qubits 4, 6, and 8 each have two neighboring qubits on which the $R_Y$ gates are applied, making them more susceptible to crosstalk effects compared to the edge qubits, which only have one such neighbor. In addition, these (physical) ions are positioned more centrally, where the spacing between ions is smaller than at the ends of the chain. Consequently, the induced crosstalk is higher in the center than at the edges, further increasing the likelihood of measuring 1 on ions 4, 6, and 8.

To investigate how the crosstalk signature depends on the rotation angle, we conducted a set of benchmark circuits. In each benchmark circuit, we applied ten consecutive $R_Y(\theta)$ gates on the same four qubits, with $\theta$ taking values from the set $\{\pi/100, \pi/8, 2\pi/8, 3\pi/8, 4\pi/8, 5\pi/8, 6\pi/8, 7\pi/8, 8\pi/8\}$. After each gate, a barrier is inserted to prevent circuit optimization by the compiler. For every angle, we perform 2000 shots and measure the five neighboring ions $[2,4,6,8,10]$ to characterize the pattern of crosstalk as a function of the applied rotation. 

We calculate the qubit-based mean squared error \begin{equation}\label{eq:mse}
\mathrm{MSE}\bigl(s_{x}^{(\mathrm{victim})},s_j^{(\mathrm{bench})}\bigr)
=
\frac{1}{n}\sum_{i=1}^{n}\bigl(y_i^{(\mathrm{victim})}-y_{ij}^{(\mathrm{bench})}\bigr)^{2},
\end{equation}
where
\begin{itemize}
  \item \(n = 5\) is the number of “listening” qubits,
  \item \(y_i^{(\mathrm{victim})}\) is the measured frequency of outcome 1 on the \(i\)-th listening qubit for the victim circuit
  \item \(y_{ij}^{(\mathrm{bench})}\) is the corresponding frequency for the \(i\)-th qubit in benchmark circuit \(j\).
\end{itemize} to compare the crosstalk signatures of the benchmark circuits with that of the victim circuit. The results are shown in the following Table \ref{tab:benchmark_mse_aqt_ibex}.

\begin{table}[h!]
    \centering
        \caption{Benchmark circuits sorted by  qubit-based MSE.}
    \label{tab:benchmark_mse_aqt_ibex}
    \begin{tabular}{lc}
        \toprule
        \textbf{Rotation Angle} & \textbf{(qubit-based) MSE} \\
        \midrule
        \rowcolor{victimcolor}
        $4\pi/8$    & 0.0000047 \\
        $3\pi/8$    & 0.0000895 \\
        $2\pi/8$    & 0.0001704 \\
        $5\pi/8$    & 0.0002433 \\
        $\pi/8$     & 0.0002739 \\
        $\pi/100$   & 0.0003223 \\
        $8\pi/8$    & 0.0003227 \\
        $6\pi/8$    & 0.0007585 \\
        $7\pi/8$    & 0.0022040 \\
        \bottomrule
    \end{tabular}
\end{table}

This result is very encouraging, as the lowest mean squared error is achieved for the benchmark circuit with rotation angle $4\pi/8 = \pi/2$, which exactly matches the angle used in the victim circuit. This shows that the attack succeeded in inferring the rotation angle solely on the basis of observed crosstalk signature. When comparing the number of observed $00000$ bitstrings in the victim circuit  with the benchmark circuits, it becomes apparent from Table~\ref{tab:zero_string_counts_vs_angle} that the benchmark circuit with rotation angle $4\pi/8 = \pi/2$ yields a count closest to the victim. This is exactly the rotation angle that was applied in the victim circuit.

\begin{table}[h!]
    \centering
        \caption{Number of occurrences of bitstring 00000 for each rotation angle.}
    \label{tab:zero_string_counts_vs_angle}
    \begin{tabular}{lc}
        \toprule
        \textbf{Rotation Angle} & \textbf{Counts of 00000} \\
        \midrule
        $\pi/100$   & 1998 \\
        $\pi/8$     & 1986 \\
        $2\pi/8$    & 1957 \\
        $3\pi/8$    & 1935 \\
        \rowcolor{victimcolor}
        $4\pi/8$    & 1839 \\
        $5\pi/8$    & 1731 \\
        $6\pi/8$    & 1660 \\
        $7\pi/8$    & 1516 \\
        $8\pi/8$    & 1489 \\
        \bottomrule
    \end{tabular}
\end{table}

The results show that the crosstalk increases with the rotation angle. Larger rotation angles induce stronger crosstalk and thus fewer $00000$ counts. 

\section{Simulated AQT Power Traces}\label{sec:simulated_aqt_power_traces}
The following plots show the simulated power traces of benchmark circuits 7, 8, and 9 discussed in Section \ref{subsec:side_channel_reconnaissance}, each compared against the power trace of the victim QNN.
\begin{figure}[htpb!]
  \centering
  \includegraphics[
      width=0.45\textwidth,
      alt={Overlay of two simulated power traces over time, comparing the victim QNN and benchmark circuit 7 on an AQT device model, showing nearly identical stepwise amplitude profiles across the full execution duration and indicating a close structural match between both circuits}
  ]{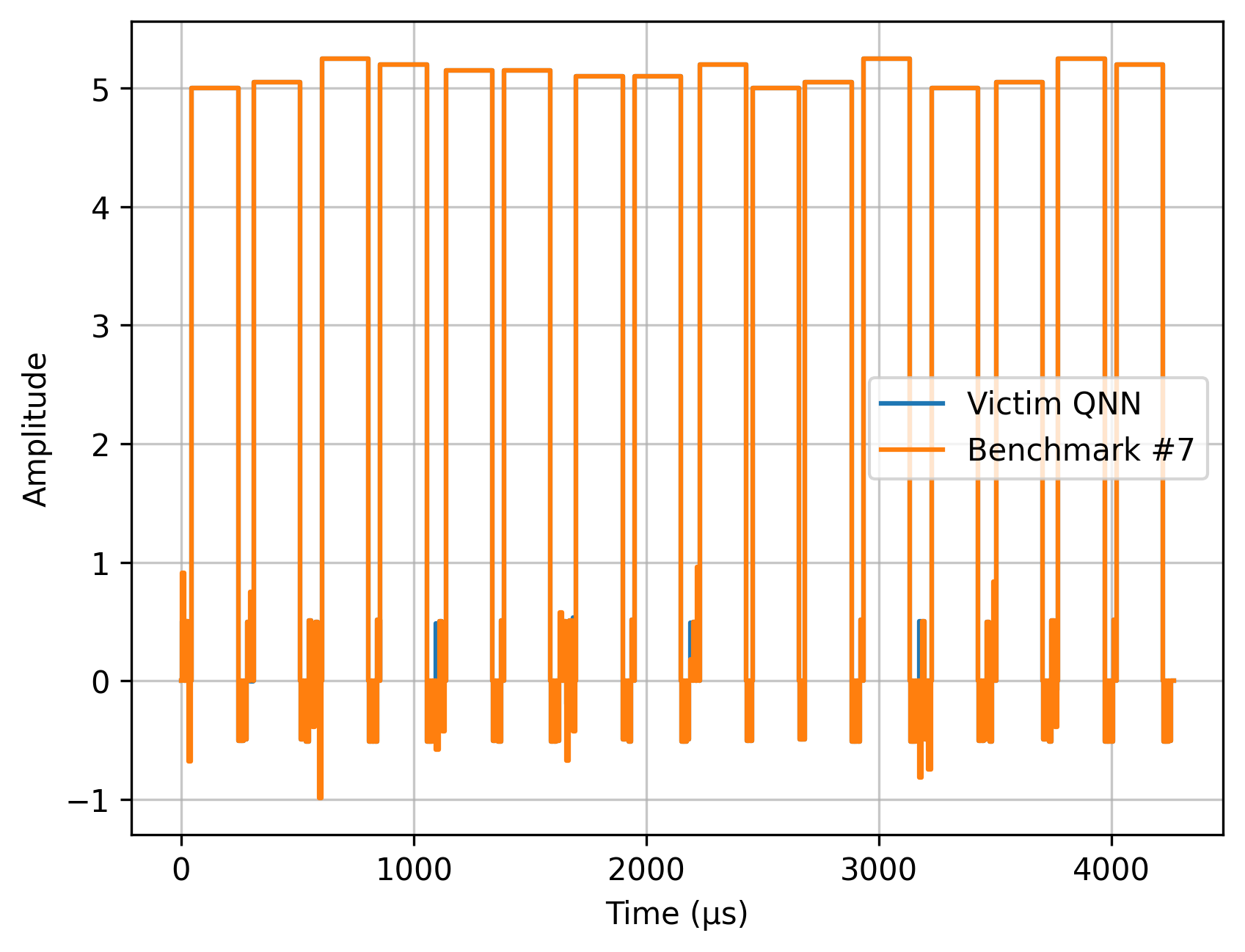}
  \caption{Power trace comparison: Victim QNN vs.\ Benchmark Circuit 7}
  \label{fig:powertrace7}
\end{figure}

\begin{figure}[htpb!]
    \centering

    \begin{minipage}[t]{0.45\textwidth}
        \centering
        \includegraphics[
            width=\textwidth,
            alt={Simulated power trace over time comparing the victim QNN with benchmark circuit 8, showing noticeable deviations in the stepwise amplitude profile and reduced alignment compared to the best matching benchmark}
        ]{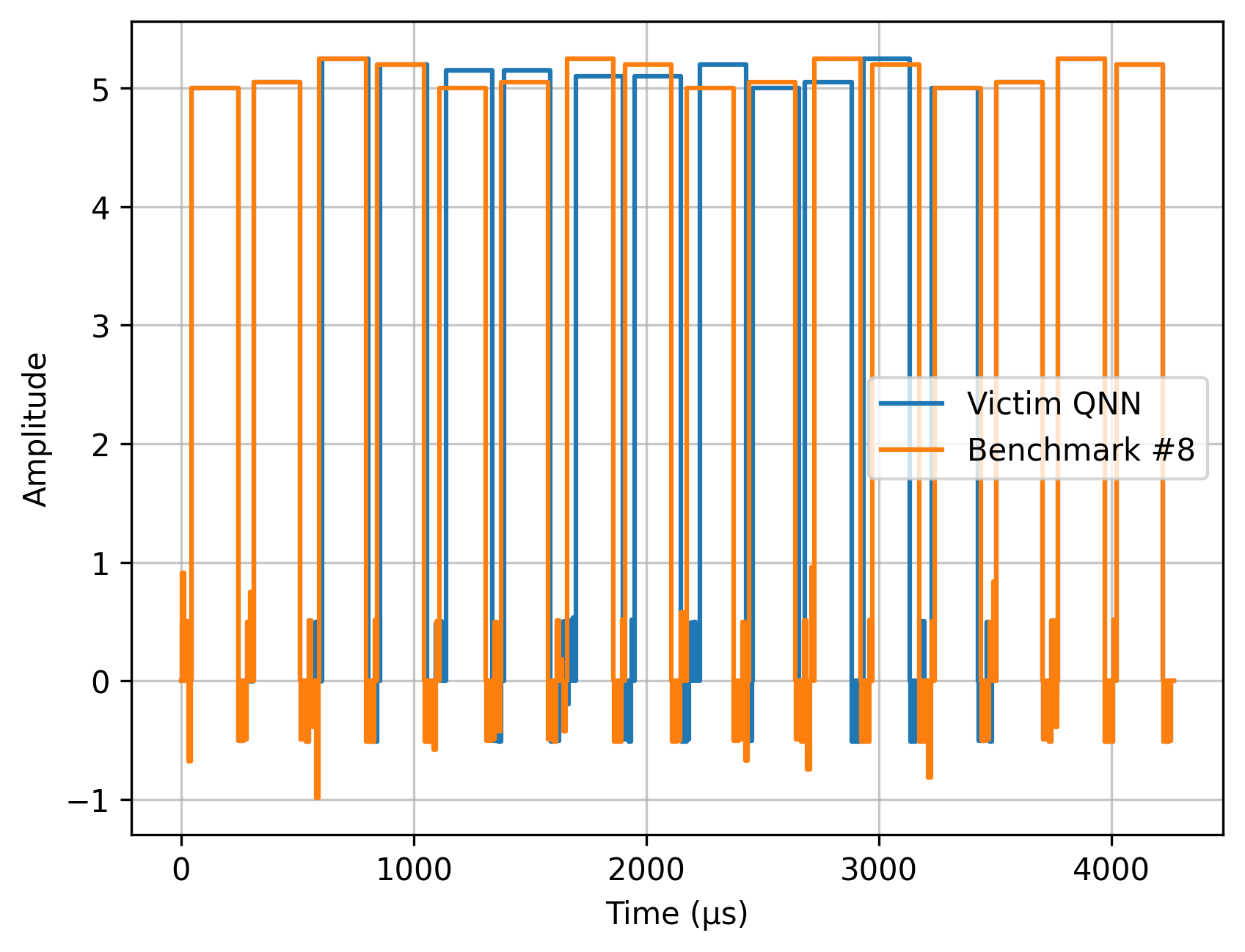}
    \end{minipage}\hfill
    \begin{minipage}[t]{0.45\textwidth}
        \centering
        \includegraphics[
            width=\textwidth,
            alt={Simulated power trace over time comparing the victim QNN with benchmark circuit 9, illustrating substantial differences in amplitude steps and timing relative to the victim trace}
        ]{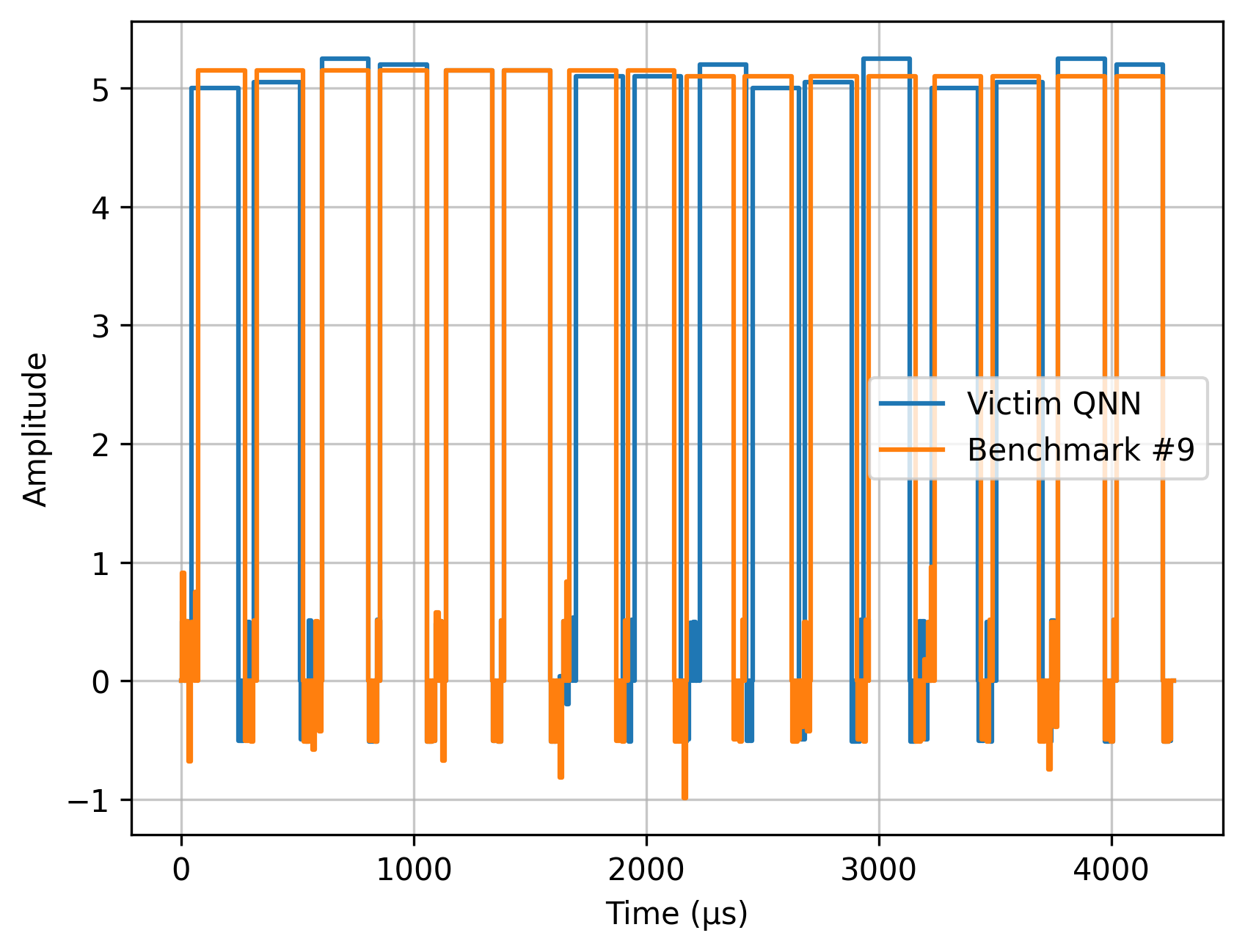}
    \end{minipage}

    \caption{Power trace comparison of the victim QNN with Benchmark Circuits 8 and 9.}
    \label{fig:powertrace8_9}
\end{figure}
\newpage 

The discrepancy between these plots mainly results from the different entangling patterns visualized in the following plots.

\begin{figure}[htbp!]
  \centering
  \includegraphics[
      width=0.9\linewidth,
      alt={Quantum circuit diagram showing the entangling structure of benchmark circuit 7 with the default setting r equal to None, illustrating the specific pattern and ordering of two qubit interactions used in the circuit}
  ]{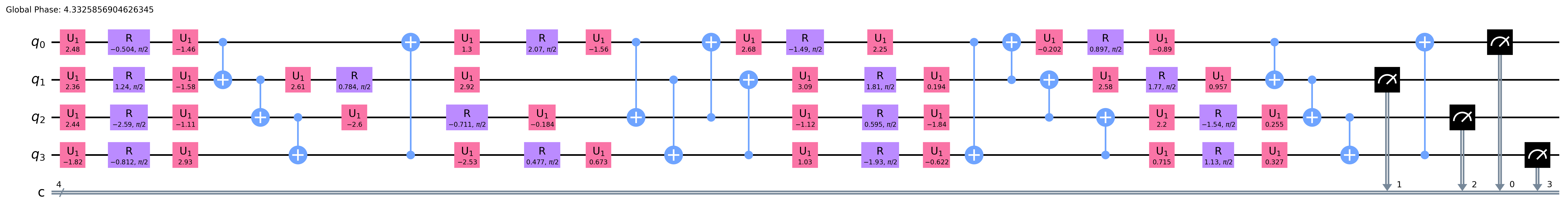}
  \caption{Benchmark Circuit 7: Entangling pattern \(r=None\).}
  \label{fig:benchmark_circuit7}
\end{figure}

\begin{figure}[htbp!]
  \centering
  \includegraphics[
      width=0.9\linewidth,
      alt={Quantum circuit diagram showing the entangling structure of benchmark circuit 8 with entanglement parameter r equal to one, highlighting a modified pattern of two qubit gates compared to the default configuration}
  ]{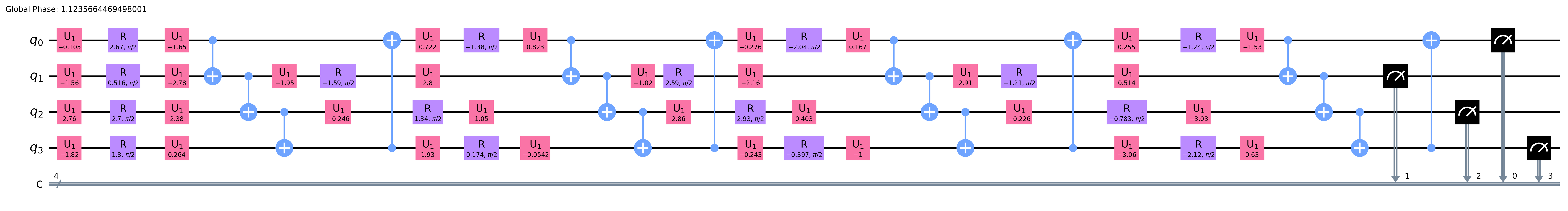}
  \caption{Benchmark Circuit 8: Entangling pattern \(r=1\).}
  \label{fig:benchmark_circuit8}
\end{figure}

\begin{figure}[htbp!]
  \centering
  \includegraphics[
      width=0.9\linewidth,
      alt={Quantum circuit diagram showing the entangling structure of benchmark circuit 9 with entanglement parameter r equal to two, depicting a denser and more extended pattern of two qubit interactions than in the lower r configurations}
  ]{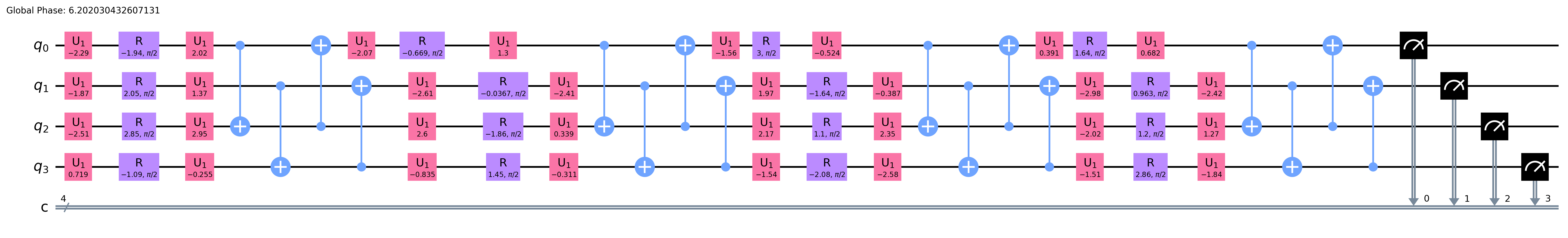}
  \caption{Benchmark Circuit 9: Entangling pattern \(r=2\).}
  \label{fig:benchmark_circuit9}
\end{figure}

\addcontentsline{toc}{section}{References}
\setlength{\bibitemsep}{0.65em}

\newpage 
{\printbibliography[title=References]}

\end{document}